\documentclass[a4paper]{article}

\usepackage[english]{babel}
\usepackage[utf8x]{inputenc}
\usepackage[T1]{fontenc}
\usepackage[margin=1in]{geometry}
\usepackage{amsmath}
\usepackage[ruled]{algorithm2e}
\usepackage{graphicx}
\usepackage{amsthm}
\usepackage{enumitem}
\usepackage{amssymb}
\usepackage[semicolon]{natbib}
\usepackage[colorinlistoftodos]{todonotes}
\usepackage[colorlinks=true, allcolors=blue]{hyperref}
\usepackage{caption}
\usepackage{graphicx}
\usepackage{listings}
\usepackage{animate}
\usepackage{multirow}
\usepackage{graphics}
\usepackage{enumitem}
\usepackage{setspace}
\usepackage{xmpmulti}
\usepackage{bbm}
\usepackage{tikz}
\usetikzlibrary{positioning}
\usetikzlibrary{shapes}
\usetikzlibrary{fit}
\usetikzlibrary{backgrounds}
\usepackage{babel}
\usepackage{titling}
\usepackage{blindtext}
\usepackage{float}
\usepackage{relsize}
\usepackage{subcaption}
\usepackage{bbm}
\usepackage[framemethod=tikz]{mdframed}
\usepackage{color}
\usepackage{ulem}
\usepackage{cleveref}

\tikzset{
  treenode/.style = {align=center, inner sep=0pt, text centered, font=\sffamily, minimum size=1.25cm},
  circnode/.style = {treenode, circle, white, font=\sffamily\bfseries, draw=black},
  circnodee/.style = {align=center, inner sep=0pt, text centered, font=\sffamily, minimum size=0.3cm, circle, white, font=\sffamily\bfseries, draw=black, fill=black},
  rectnode/.style = {treenode, rectangle, draw=black, minimum size=1em},
  xnode/.style = {treenode, cross out, white, font=\sffamily\bfseries, draw=red, fill=black, ultra thick}
}

\newcommand{\beginsupplement}{%
        \setcounter{table}{0}
        \renewcommand{\thetable}{S\arabic{table}}%
        \setcounter{figure}{0}
        \renewcommand{\thefigure}{S\arabic{figure}}%
     }

\allowdisplaybreaks

\newcommand{\Categorical}{\operatorname{Categorical}}
\newcommand{\Exponential}{\operatorname{Exponential}}
\newcommand{\Beta}{\operatorname{Beta}}
\newcommand{\Uniform}{\operatorname{Uniform}}
\newcommand{\GGamma}{\operatorname{Gamma}}

\title{A Bayesian Phylogenetic Hidden Markov Model for B Cell Receptor Sequence Analysis}

\author{
  Amrit Dhar$^{1,2}$, Duncan K. Ralph$^{2}$, Vladimir N. Minin$^{3,*}$, Frederick A. Matsen IV$^{2,*}$\\ \\
  $^1$Department of Statistics, University of Washington, Seattle \\
  $^2$Fred Hutchinson Cancer Research Center\\
  $^3$Department of Statistics, University of California, Irvine \\
  $^{*}$corresponding authors: \url{vminin@uci.edu}, \url{matsen@fredhutch.org}
}

\date{\today}

\begin{document}
\maketitle

\begin{abstract}
The human body is able to generate a diverse set of high affinity antibodies, the soluble form of B cell receptors (BCRs), that bind to and neutralize invading pathogens.
The natural development of BCRs must be understood in order to design vaccines for highly mutable pathogens such as influenza and HIV.
BCR diversity is induced by naturally occurring combinatorial ``V(D)J'' rearrangement, mutation, and selection processes.
Most current methods for BCR sequence analysis focus on separately modeling the above processes.
Statistical phylogenetic methods are often used to model the mutational dynamics of BCR sequence data, but these techniques do not consider all the complexities associated with B cell diversification such as the V(D)J rearrangement process.
In particular, standard phylogenetic approaches assume the DNA bases of the progenitor (or ``naive'') sequence arise independently and according to the same distribution, ignoring the complexities of V(D)J rearrangement.
In this paper, we introduce a novel approach to Bayesian phylogenetic inference for BCR sequences that is based on a phylogenetic hidden Markov model (phylo-HMM).
This technique not only integrates a naive rearrangement model with a phylogenetic model for BCR sequence evolution but also naturally accounts for uncertainty in all unobserved variables, including the phylogenetic tree, via posterior distribution sampling.
\end{abstract}

\section*{Introduction}

One of the most important features of the adaptive immune system is its ability to create a wide variety of high affinity antibodies, the soluble form of B cell receptors (BCRs), that bind to and neutralize pathogens in the body.
The initial BCR diversity is generated by randomly joining together various gene segments in a process called V(D)J rearrangement; after an initial testing process the cells reach the ``naive'' state.
When stimulated by binding to foreign material called ``antigen,'' B cells diversify further by entering germinal centers (GCs) in the secondary lymphoid organs and going through an affinity maturation process.
During the GC reaction, B cells mutate rapidly in a process called somatic hypermutation (SHM), and the high affinity clones are positively selected for via clonal expansion.
We would like to better understand the GC mutation and selection processes, because insight into mutational pathways from naive to mature BCR sequences could aid in the development of vaccines for highly mutable pathogens such as influenza and HIV \citep{Mascola2013-mt}.
We have developed a new statistical inference framework that better estimates these mutational pathways and quantifies uncertainty in these estimates.

Rational vaccine design efforts depend on accurate inference of full evolutionary paths from a given naive sequence to the corresponding mature BCR sequences in a GC.
By understanding the mutational pathways that lead to broadly neutralizing antibodies (bNAbs), vaccines could then be constructed that induce the production of these bNAbs in the body \citep{Stamatatos2017-ut}.
For instance, in the case of HIV, most bNAbs are generally not observed until after a long period of chronic infection, requiring prospective studies to characterize them \citep{liao2013co}.
While there have been many such studies that experimentally test longitudinal B cell samples spanning from an initial HIV infection to the development of mature bNAbs \citep{liao2013co,doria2014developmental,doria2016new}, obtaining early prospective samples is difficult.
One way to avoid the process described above is to infer intermediate lineage sequences computationally, synthesize them in the laboratory, and test their binding and neutralization abilities \citep{doria2014developmental,simonich2019kappa}.
Consequently, we will focus on inferring mutational pathways through ancestral sequence reconstruction, a commonly used technique in computational phylogenetics.

Much of the existing BCR sequence analysis literature focuses on modeling either the V(D)J recombination process, or the phylogenetic diversification process, but not both.
\citet{elhanati2015inferring} develop a likelihood-based model that encompasses the V(D)J rearrangement, SHM, and clonal selection processes; however, they implicitly assume clonal sequences arise independently and do not consider the phylogenetic structure of SHM.
\citet{Hoehn417,Hoehn2019-jo} introduce novel codon substitution models for ancestral lineage inference that encodes SHM context-dependent mutational effects but does not account for V(D)J rearrangement dynamics in the naive sequence within their maximum likelihood phylogenetic inference framework.
\citet{yaari2013models} provide estimates of context-dependent SHM substitution probabilities and motif mutability scores based on an aggregated dataset consisting of the synonymous codon positions of productive antibody sequences.
\citet{ralph2016consistency} are able to infer naive BCR sequences using an HMM-based approach that models the V(D)J rearrangement process but assumes independent evolution across the different lineages.
While these efforts have contributed greatly to our understanding of GC dynamics, we believe that the performance of clonal lineage and ancestral sequence inference procedures can be enhanced by using an evolutionary model for SHM that also accounts for uncertainty in the naive rearrangement process.

\citet{kepler2013reconstructing} has developed a likelihood-based SHM modeling framework that jointly estimates the naive sequence and the associated clonal tree and incorporates information about the V(D)J rearrangement process.
However, this work does not consider phylogenetic or ancestral sequence uncertainty in the naive sequence estimation procedure.
While there is evidence to suggest that ancestral sequence estimation is robust to phylogenetic uncertainty in other settings \citep{hanson2010robustness}, the parameter regime of BCR diversification is quite different from that of this previous work, leaving open the question of whether incorporating phylogenetic uncertainty would aid in ancestral sequence inference.
Therefore, we would like to construct a phylogenetic inference procedure that not only allows for easy quantification of phylogenetic and ancestral sequence uncertainty but also models the V(D)J recombination as an informative prior for the naive sequence at the root of a phylogenetic tree describing the evolution of one clonal lineage.

In this paper, we propose a Bayesian approach to phylogenetic inference for clonal sequences that is based on a phylogenetic hidden Markov model (phylo-HMM) \citep{siepel2005phylogenetic}.
Our phylo-HMM models both the naive rearrangement and SHM processes. The Bayesian framework allows us to naturally account for uncertainty in all unobserved variables, including a phylogenetic tree, via posterior distribution sampling.
We perform simulation-based experiments to show that naive sequence and phylogenetic inference performed jointly provides higher-quality estimates than those obtained by considering these inferences separately.
Our application to real data reveals significant uncertainty in naive and ancestral sequences, confirming the importance of a Bayesian approach.

\section*{Methods}

\subsection*{Overview}
\begin{figure}[!ht]
{\Large (A)}
\vfill
\begin{subfigure}{\textwidth}
\centerline{\includegraphics[width=0.7\textwidth]{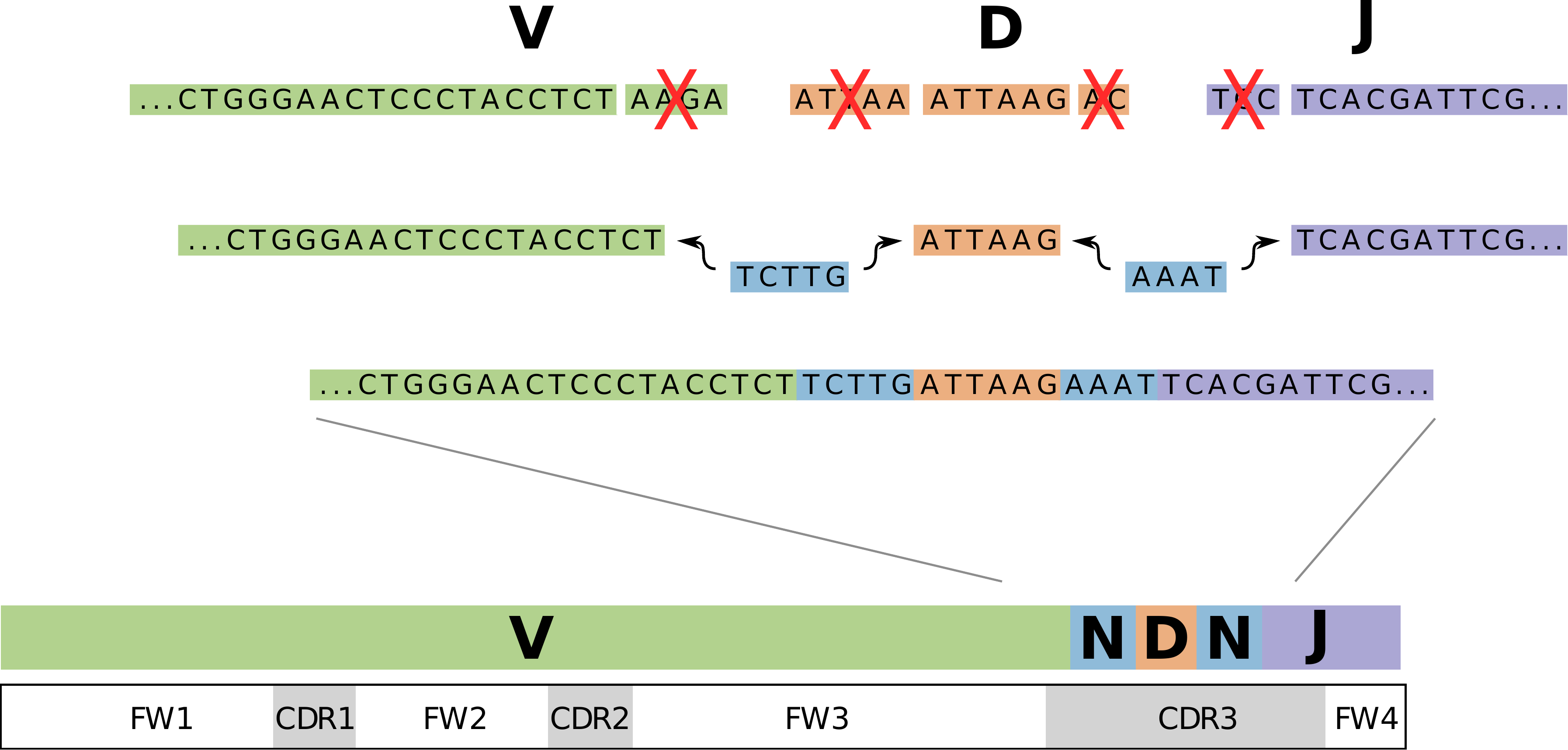}}
\phantomsubcaption{}
\label{vdj_recomb}
\end{subfigure}
\newline
\vspace{5mm}
\newline
{\Large (B)}
\vfill
\begin{subfigure}{\textwidth}
\centering
\includegraphics[scale=0.6]{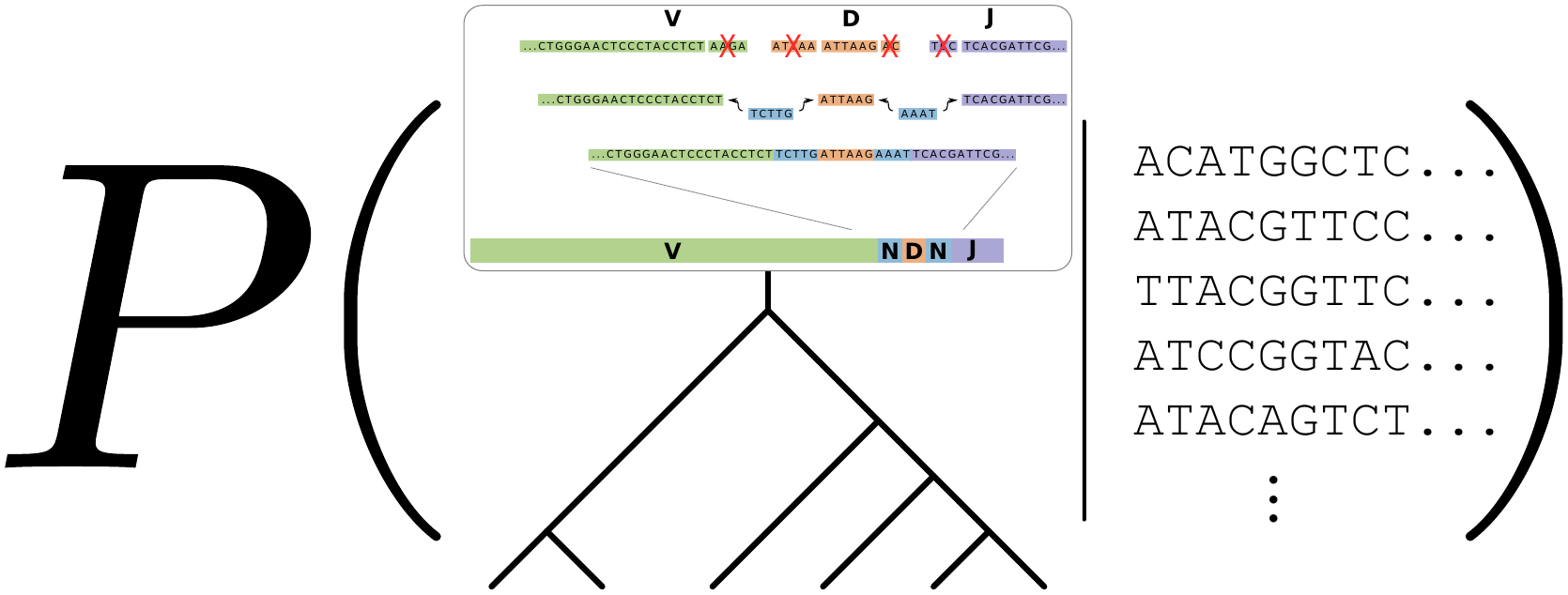}
\phantomsubcaption{}
\label{phylo-hmm}
\end{subfigure}
\caption{(A) A schematic representation of the naive rearrangement process.  First, V (green), D (orange), and J (purple) genes are randomly selected from the respective gene pools in the body.  Then, nucleotides are randomly deleted (red X's) from both ends of the V-D and D-J junction regions and random bases (blue) are added to the same junction regions before the V, D, and J germline genes can be joined together.  The BCR sequences can be partitioned into framework (FWK) and complementarity-determining (CDR) regions.  This image was taken from \citep{ralph2016consistency}.  (B) Our Bayesian phylo-HMM jointly models V(D)J recombination at the root of the tree (using an HMM) and then subsequent diversification (via a phylogenetic tree).  We do posterior inference conditioning on the observed sequence alignment in a clonal family, but not on a fixed inferred naive sequence.}
\end{figure}

The immune system is able to generate a diverse set of naive BCR sequences due to the V(D)J rearrangement process (\autoref*{vdj_recomb}).
In this process, the B cell first randomly selects V, D, and J gene segments (i.e.\ DNA sequences) from the respective gene pools in the body.
Before joining the gene segments together, the B cell randomly deletes nucleotides at both ends of the V-D and D-J junction regions and randomly inserts nucleotides in the same junction regions (i.e.\ non-templated insertions).
BCRs comprise both heavy chain and light chain sequences, where the former goes through the VDJ rearrangement process described above while the latter undergoes VJ recombination, a process similar to VDJ recombination without D germline genes; in this paper, we focus solely on heavy chain inference, but our methods and implementation extend to light chain inference too.
Although the V(D)J rearrangement process samples germline genes from the same gene pools for all the naive BCRs in a given individual, different people may have different collections of germline genes \citep{Watson2017-mr}.
BCR sequences can be partitioned into framework (FWK) and complementarity-determining (CDR) regions.
The BCR binding affinity is largely determined by the sequence segments in the CDR regions, and among all the CDR regions the CDR3 region contributes the most to antigen-binding specificity and has the highest amount of sequence variability.
The \texttt{partis} software program \citep{ralph2016consistency, ralph2016likelihood} treats the naive sequence generative process as a discrete-time Markov chain (DTMC) going from left to right across the sequence bases and also permits maximum likelihood inference of the associated model parameters, which are defined according to the V(D)J rearrangement dynamics.
We emphasize that the ``time'' of this DTMC represents position along the sequence, in contrast to the continuous time Markov chain described below modeling the substitution process through chronological time.

As we attempt to characterize the affinity maturation process in GCs, we use the concept of a clonal family (CF) to help analyze BCR repertoire datasets that result from high-throughput sequencing experiments.
A clonal family represents a set of BCR sequences that originate from the same naive rearrangement event; in practice, it is simpler to define these families as groups of BCR sequences that share the same naive sequence \citep{ralph2016likelihood}.
In this work, we use the latter definition of a CF.
The CF definition used here relies on the basic assumption that the unmutated common ancestor of a collection of clonally related sequences can only be identified by the corresponding DNA sequence.
There is a chance that two different naive B cells with identical BCR sequences could seed multiple GCs and form their own lineages, but the observed clones from the two GCs would be collapsed into a single CF because both lineages share the same naive sequence.
Thus, a correctly inferred CF under this definition will be a cluster of sequences that derive from naive B cells with the same BCR sequence; because GC mutation and selection occur at the sequence level, all sequences in a CF go through the same mutation and selection processes.
Using this CF definition, \citet{ralph2016likelihood} describe how to cluster BCR sequences from large-scale repertoire datasets into CFs with high accuracy using \texttt{partis}.
We want to emphasize that repertoire datasets can be clustered and pre-processed in many different ways, but \texttt{partis} provides a convenient way to cluster repertoire sequences and produce CF-specific multiple sequence alignments (MSAs).

\begin{figure}[!ht]
\centering
\begin{tikzpicture}
  \node[circnodee, label={[label distance=0.0325cm]45:$x$}] (x) {};
  \node[circnodee, label={[label distance=0.0325cm]90:$w$}] (w) [above=1.5cm of x] {};
  \node[circnodee, label={[label distance=0.0325cm]90:$y$}] (y) [below left=1cm and 2cm of x] {};
  \node[circnodee, label={[label distance=0.0325cm]90:$z$}] (z) [below right=1cm and 2cm of x] {};
  \node[circnodee, label={[label distance=0.1cm]270:A}] (term1) [below left=2.5cm and 1cm of y] {};
  \node[circnodee, label={[label distance=0.1cm]270:T}] (term2) [below right=2.5cm and 1cm of y] {};
  \node[circnodee, label={[label distance=0.1cm]270:C}] (term3) [below left=2.5cm and 1cm of z] {};
  \node[circnodee, label={[label distance=0.1cm]270:T}] (term4) [below right=2.5cm and 1cm of z] {};
  \path[ultra thick] (w) edge node[left] {$t_0$} (x);
  \path[ultra thick] (x) edge node[above left] {$t_1$} (y);
  \path[ultra thick] (x) edge node[above right] {$t_2$} (z);
  \path[ultra thick] (y) edge node[above left] {$t_3$} (term1);
  \path[ultra thick] (y) edge node[above right] {$t_4$} (term2);
  \path[ultra thick] (z) edge node[above left] {$t_5$} (term3);
  \path[ultra thick] (z) edge node[above right] {$t_6$} (term4);
\end{tikzpicture}
\caption{An example phylogenetic tree.  Letters $x,y,z,w$ represent the unobserved internal node states where $w$ is associated with the root node, $t_0$ defines the root branch length, and $(t_0, t_1, t_2,..., t_6)$ denotes the entire vector of branch lengths.  Given this tree topology and set of branch lengths, we can calculate the likelihood of observing the nucleotide vector $(A,T,C,T)$ by marginalizing probabilities over the unobserved states $x,y,z,w$.}
\label{extree}
\end{figure}
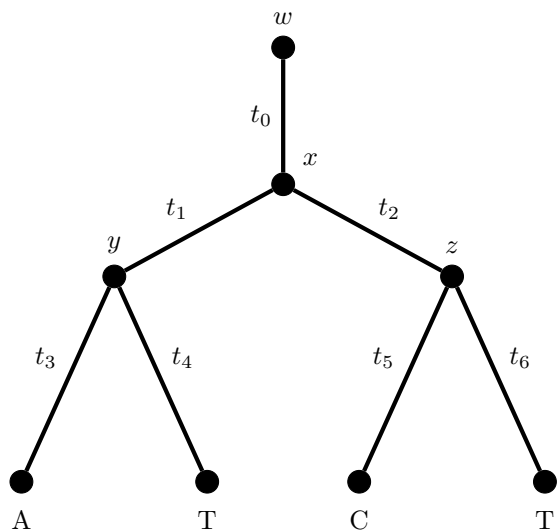

As mentioned in the opening section, naive sequences accumulate mutations via SHM and the corresponding B cells undergo cellular replication.
Phylogenetic tree models provide a realistic and mathematically convenient way to represent the aforementioned B cell evolutionary dynamics of a CF.
In particular, these models define a likelihood at each MSA position/site as a function of unknown parameters.
These parameters consist of a tree topology, branch lengths, and continuous-time Markov chain (CTMC) substitution model parameters.
While B cell evolution in a CF is not independent across the different site positions, site-specific phylogenetic likelihoods provide a convenient first approximation in modeling this phenomenon.
To help us illustrate how these phylogenetic models work, we provide an example tree visualization for 4 sequences (\autoref*{extree}).

Given a tree topology with branch lengths, we use a CTMC substitution model to calculate the probabilities of state changes along the branches of the tree.
Specifically if $t$ denotes a branch length on a tree, CTMC substitution models allow one to calculate $p_{ij}(t)$, which denotes the probability of going from state $i$ to state $j$ on a branch of length $t$, where $i,j \in \{ A,G,C,T \}$.
It is common to use a reversible CTMC substitution model on a tree \citep{felsenstein2004inferring}; a reversible substitution model is a Markov substitution model that, if started at stationarity, can be run backwards in time, with the resulting backward Markov model following the same probability law as the original forward model.
The standard phylogenetic generative process can be described at each alignment site as follows: 1) a DNA state at the root node is drawn independently according to the same 4-state discrete distribution and 2) the states at the other nodes are sampled in a pre-order traversal using the computed CTMC probabilities at each branch $p_{ij}(t)$; this model is a special case of a directed graphical model \citep{lauritzen1996graphical} that probabilistically generates sequence alignments.

Clearly, standard phylogenetic models do not account for naive rearrangement dynamics at the root because, as we discussed above, the root state at each sequence position is sampled independently according to an identical distribution.
Instead, if we draw root sequence states from the DTMC mentioned earlier, we would obtain a sequence evolution model that more accurately describes B cell evolutionary dynamics.
Thus, we formulate our phylo-HMM to consist of a hidden state DTMC model for naive sequences that explicitly incorporates V(D)J rearrangement information, and an emission distribution that generates sequence alignments conditional on the naive sequence that is based on phylogenetic likelihoods.
This phylo-HMM hopefully leads to more accurate naive sequence estimates and, as a result of that, higher-quality intermediate ancestral sequence estimates.
We introduce a pictorial representation of our Bayesian phylo-HMM to make clear our target of inference (\autoref*{phylo-hmm}).

\subsection*{Notation and Assumptions}

We now introduce some notation and assumptions that will be used throughout this paper.
Let $\mathbf{D} = \{ D_i^{(j)} \}_{i=1:m,j=1:n}$ denote the MSA of $m$ clonal DNA sequences of length $n$.
We define $\mathbf{Y}_{\text{naive}} = \{ Y_{\text{naive}}^{(j)} \}_{j=1:n}$ and $\widehat{\mathbf{Y}}_{\text{naive}} = \{ \widehat{Y}_{\text{naive}}^{(j)} \}_{j=1:n}$ to be the corresponding length-$n$ naive sequence random variable and point estimate, respectively.
We let $\tau$ represent a tree topology with $m$ tips and a root branch length; in total, this topology has $m$ internal nodes and $2m-1$ branch lengths.
We assume that the ancestral sequence at the root of $\tau$ is $\mathbf{Y}_{\text{naive}}$.
Furthermore, we define $\mathbf{t} = \{ t_i \}_{i=1:(2m-1)}$ to be the branch lengths associated with $\tau$.
Let $\mathbf{Y}_{\text{int}} = \{ Y_i^{(j)} \}_{i=1:(m-1),j=1:n}$ denote the internal nodes of $\tau$ excluding the naive sequence $\mathbf{Y}_{\text{naive}}$.
For convenience, we let $\mathbf{D}^{(j)} = \{ D_i^{(j)} \}_{i=1:m}$ and $\mathbf{Y}_{\text{int}}^{(j)} = \{ Y_i^{(j)} \}_{i=1:(m-1)}$ symbolize the observed sequence data and unobserved ancestral sequence data, respectively, at MSA site $j \in \{ 1, ..., n \}$.
Conditioned on the root sequence $\mathbf{Y}_{\text{naive}}$, we assume that the ancestral states at each site in the MSA evolve independently along the phylogeny $\tau$ according to a general time-reversible (GTR) substitution model \citep{tavare1986some}.
Let $\mathbf{e} = \{ e_{AC}, e_{AG}, e_{AT}, e_{CG}, e_{CT}, e_{GT} \}$ and $\boldsymbol{\pi} = \{ \pi_A, \pi_C, \pi_G, \pi_T \}$ represent the GTR exchangeability rates and equilibrium base frequencies, respectively.
We also account for phylogenetic rate variation among sites by employing a discrete gamma distribution with a fixed number of rate classes $K$ \citep{yang1994maximum, yang1996among} and define $\alpha$ to be the associated gamma shape parameter, denote $\mathbf{r} = \{ r_1(\alpha), ..., r_K(\alpha) \}$ as the set of discrete rates deterministically induced by $\alpha$, and let $\mathbf{r}^* = \{ r^*_{(j)} \}_{j=1:n}$ represent the discrete rates chosen at each site in the MSA.
In theory, we would like to compute phylogenetic likelihoods with branch lengths scaled by $r^*_{(j)}$ and mix over $r^*_{(j)} \sim \GGamma(\alpha, \alpha)$ for $j \in \{ 1, ..., n \}$.
However, these integrals are generally intractable so \citet{yang1994maximum} suggested dividing the $\GGamma(\alpha, \alpha)$ distribution into $K$ equal-probability rate classes, with the mean rate in each class used to represent all rates in that class.
In practice, we use the $\Categorical(\mathbf{r}, \mathbf{p})$ distribution to define the models on $r^*_{(j)}$ for $j \in \{ 1, ..., n \}$, where $\mathbf{p}$ is a, possibly unnormalized, probability vector; in addition, this model is used to represent more general discrete distributions.

From a statistical point of view, it is common to assume the naive sequence root node is a leaf node holding naive sequence bases connected to a ``virtual root node'' (i.e.\ what we call ``root'' node for our phylogenetic model described in the previous subsection) via a branch length of 0.
Even though it may seem like we have described a rooted tree model above, it turns out that under the assumptions of a reversible substitution model and a nucleotide distribution at the virtual root starting at stationarity, the Pulley Principle, first discussed in \citep{felsenstein1981evolutionary}, states that the virtual root may be placed anywhere on the tree without affecting the likelihood.
This implies that the model described above does not correspond to a single rooted tree, but an equivalence class of rooted trees that maps to a unique unrooted tree.
This is an important distinction as our phylo-HMM will in fact use an unrooted tree model, which will be justified when we describe our posterior sampling approach.

\subsection*{Phylo-HMM Description}

Phylo-HMMs are special cases of directed graphical models \citep{lauritzen1996graphical} and treat evolution as a combination of two Markov processes: one across the sites in the MSA and one down the phylogeny.
They are commonly used for sequence-level segmentation problems such as gene prediction and detection of highly-conserved regions \citep{siepel2005phylogenetic}.
In fact, a phylo-HMM is similar in structure to a standard HMM; the main difference between the two model classes is that a phylo-HMM uses a phylogenetic likelihood as its emission probability distribution, while standard HMMs usually specify simpler emission distributions.
Our BCR-specific phylo-HMM specifies a Markov process along $\mathbf{Y}_{\text{naive}}$ and, conditional on $\mathbf{Y}_{\text{naive}}$, a phylogenetic evolutionary process down the given tree.
Phylo-HMMs have not been applied to BCR sequence analysis before and we believe this biologically realistic probabilistic model is uniquely suited to provide higher-quality naive sequence and ancestral sequence estimates compared to those obtained under current state-of-the-art methods.

\begin{figure}
\centering
\resizebox{!}{1.15\textwidth}{
\begin{tikzpicture}[>=stealth]
  \node[circnode, label={[shift={(-7ex,6ex)}]south east:{\large $Y_{\text{naive}}^{(1)}$}}] (i1) {};
  \node[circnode, label={[shift={(-7ex,6ex)}]south east:{\Large $Y_1^{(1)}$}}] (i2) [below=1cm of i1] {};
  \node[circnode, label={[shift={(-7ex,6ex)}]south east:{\Large $Y_2^{(1)}$}}] (i4) [below left=1.25cm and 0.1cm of i2] {};
  \node[circnode, label={[shift={(-7ex,6ex)}]south east:{\Large $D_3^{(1)}$}}] (d3) [below right=3.15cm and 0.5cm of i2] {};
  \node[circnode, label={[shift={(-7ex,6ex)}]south east:{\Large $D_1^{(1)}$}}] (d1) [below left=1cm and 0.005cm of i4] {};
  \node[circnode, label={[shift={(-7ex,6ex)}]south east:{\Large $D_2^{(1)}$}}] (d2) [below right=1cm and 0.005cm of i4] {};
  \path[very thick, ->] (i1) edge (i2);
  \path[very thick, ->] (i2) edge (i4);
  \path[very thick, ->] (i2) edge (d3);
  \path[very thick, ->] (i4) edge (d1);
  \path[very thick, ->] (i4) edge (d2);
  \node[circnode, label={[shift={(-7ex,6ex)}]south east:{\large $Y_{\text{naive}}^{(2)}$}}] (i1_2) [right=4cm of i1] {};
  \path[very thick, ->] (i1) edge (i1_2);
  \node[circnode, label={[shift={(-7ex,6ex)}]south east:{\Large $Y_1^{(2)}$}}] (i2_2) [below=1cm of i1_2] {};
  \node[circnode, label={[shift={(-7ex,6ex)}]south east:{\Large $Y_2^{(2)}$}}] (i4_2) [below left=1.25cm and 0.1cm of i2_2] {};
  \node[circnode, label={[shift={(-7ex,6ex)}]south east:{\Large $D_3^{(2)}$}}] (d3_2) [below right=3.15cm and 0.5cm of i2_2] {};
  \node[circnode, label={[shift={(-7ex,6ex)}]south east:{\Large $D_1^{(2)}$}}] (d1_2) [below left=1cm and 0.005cm of i4_2] {};
  \node[circnode, label={[shift={(-7ex,6ex)}]south east:{\Large $D_2^{(2)}$}}] (d2_2) [below right=1cm and 0.005cm of i4_2] {};
  \path[very thick, ->] (i1_2) edge (i2_2);
  \path[very thick, ->] (i2_2) edge (i4_2);
  \path[very thick, ->] (i2_2) edge (d3_2);
  \path[very thick, ->] (i4_2) edge (d1_2);
  \path[very thick, ->] (i4_2) edge (d2_2);
  \node[circnode, label={[shift={(-7ex,6ex)}]south east:{\large $Y_{\text{naive}}^{(3)}$}}] (i1_3) [right=4cm of i1_2] {};
  \path[very thick, ->] (i1_2) edge (i1_3);
  \node[circnode, label={[shift={(-7ex,6ex)}]south east:{\Large $Y_1^{(3)}$}}] (i2_3) [below=1cm of i1_3] {};
  \node[circnode, label={[shift={(-7ex,6ex)}]south east:{\Large $Y_2^{(3)}$}}] (i4_3) [below left=1.25cm and 0.1cm of i2_3] {};
  \node[circnode, label={[shift={(-7ex,6ex)}]south east:{\Large $D_3^{(3)}$}}] (d3_3) [below right=3.15cm and 0.5cm of i2_3] {};
  \node[circnode, label={[shift={(-7ex,6ex)}]south east:{\Large $D_1^{(3)}$}}] (d1_3) [below left=1cm and 0.005cm of i4_3] {};
  \node[circnode, label={[shift={(-7ex,6ex)}]south east:{\Large $D_2^{(3)}$}}] (d2_3) [below right=1cm and 0.005cm of i4_3] {};
  \path[very thick, ->] (i1_3) edge (i2_3);
  \path[very thick, ->] (i2_3) edge (i4_3);
  \path[very thick, ->] (i2_3) edge (d3_3);
  \path[very thick, ->] (i4_3) edge (d1_3);
  \path[very thick, ->] (i4_3) edge (d2_3);
  \node[circnode, label={[shift={(-4.5ex,5ex)}]south east:{\Large $\mathbf{t}$}}, label={:{element-wise}}] (t) [above left=4cm and 0.25cm of i1_2]{};
  \node[draw,inner sep=2mm,fit=(i2) (i4) (d1) (d2) (d3)] (site1) {};
  \path[very thick, ->] (t) edge (site1);
  \node[draw,inner sep=2mm,fit=(i2_2) (i4_2) (d1_2) (d2_2) (d3_2)] (site2) {};
  \path[very thick, ->] (t) edge[bend right=10] (site2);
  \node[draw,inner sep=2mm,fit=(i2_3) (i4_3) (d1_3) (d2_3) (d3_3)] (site3) {};
  \path[very thick, ->] (t) edge (site3);
  \node[circnode, label={[shift={(-5ex,4.5ex)}]south east:{\Large $\tau$}}] (tau) [left=1cm of t]{};
  \path[very thick, ->] (tau) edge (site1);
  \path[very thick, ->] (tau) edge (site2);
  \path[very thick, ->] (tau) edge[bend left=15] (site3);
  \node[circnode, label={[shift={(-5ex,4.5ex)}]south east:{\Large $\boldsymbol{\pi}$}}] (pi) [above right=4cm and 0.25cm of i1_2]{};
  \node[circnode, label={[shift={(-5ex,4.5ex)}]south east:{\Large $\mathbf{e}$}}] (e) [right=1cm of pi]{};
  \path[very thick, ->] (pi) edge (site1);
  \path[very thick, ->] (pi) edge[bend left=15] (site2);
  \path[very thick, ->] (pi) edge (site3);
  \path[very thick, ->] (e) edge (site1);
  \path[very thick, ->] (e) edge[bend left=10] (site2);
  \path[very thick, ->] (e) edge (site3);
  \node[circnode, label={[shift={(-6ex,6ex)}]south east:{\Large $r_{(1)}^{*}$}}] (r_1) [below=1.5cm of site1]{};
  \node[circnode, label={[shift={(-6ex,6ex)}]south east:{\Large $r_{(2)}^{*}$}}] (r_2) [below=1.5cm of site2]{};
  \node[circnode, label={[shift={(-6ex,6ex)}]south east:{\Large $r_{(3)}^{*}$}}] (r_3) [below=1.5cm of site3]{};
  \path[very thick, ->] (r_1) edge (site1);
  \path[very thick, ->] (r_2) edge (site2);
  \path[very thick, ->] (r_3) edge (site3);
  \node[circnode, label={[shift={(-4.625ex,4.5ex)}]south east:{\Large $\mathbf{r}$}}] (r) [below=1cm of r_2]{};
  \path[very thick, ->] (r) edge (r_1);
  \path[very thick, ->] (r) edge (r_2);
  \path[very thick, ->] (r) edge (r_3);
  \node[circnode, label={[shift={(-5ex,4.5ex)}]south east:{\Large $\alpha$}}] (alpha) [below=1.25cm of r]{};
  \path[very thick, ->] (alpha) edge (r);
\end{tikzpicture}
}
\caption{The phylo-HMM graphical model diagram for an example alignment with $m=3$ sequences and $n=3$ sites.  The $\tau$, $\mathbf{t}$, $\boldsymbol{\pi}$, and $\mathbf{e}$ nodes represent the $4$-tip unrooted tree topology, the associated $5$ branch lengths, the GTR exchangeability rates, and GTR equilibrium base frequencies, respectively.  The parameter $\alpha$ denotes the gamma shape parameter associated with the $K$-class discrete gamma distribution, which is used to model phylogenetic rate variation among sites; $\mathbf{r}$ symbolizes the vector of $K$ discrete rates that is deterministically induced by $\alpha$.  The set of nodes $\mathbf{r}^* = \{ r^*_{(1)}, r^*_{(2)}, r^*_{(3)} \}$ defines the rates that are drawn from $\mathbf{r}$ at each particular site.  The $\mathbf{Y}_{\text{naive}} = \{ Y_{\text{naive}}^{(1)}, Y_{\text{naive}}^{(2)}, Y_{\text{naive}}^{(3)} \}$ ``hidden state'' node collection represents the Markov process that stochastically generates the naive sequence in our phylo-HMM.  The node sets $\{ Y_i^{(j)} \}_{i=1:2,j=1:3}$ and $\mathbf{D} = \{ D_i^{(j)} \}_{i=1:3,j=1:3}$ denote the internal nodes of $\tau$ excluding the naive sequence $\mathbf{Y}_{\text{naive}}$ and the observed MSA, respectively.  We draw plates around the $\mathbf{Y}_{\text{int}}^{(j)}$ and $\mathbf{D}^{(j)}$ node sets for $j \in \{ 1, 2, 3 \}$ to indicate that any directed edges touching a plate apply to all nodes in the plate (except for edges that originate from $\mathbf{t}$, which apply element-wise to the nodes in the plate).}
\label{phylo-hmm-dag}
\end{figure}
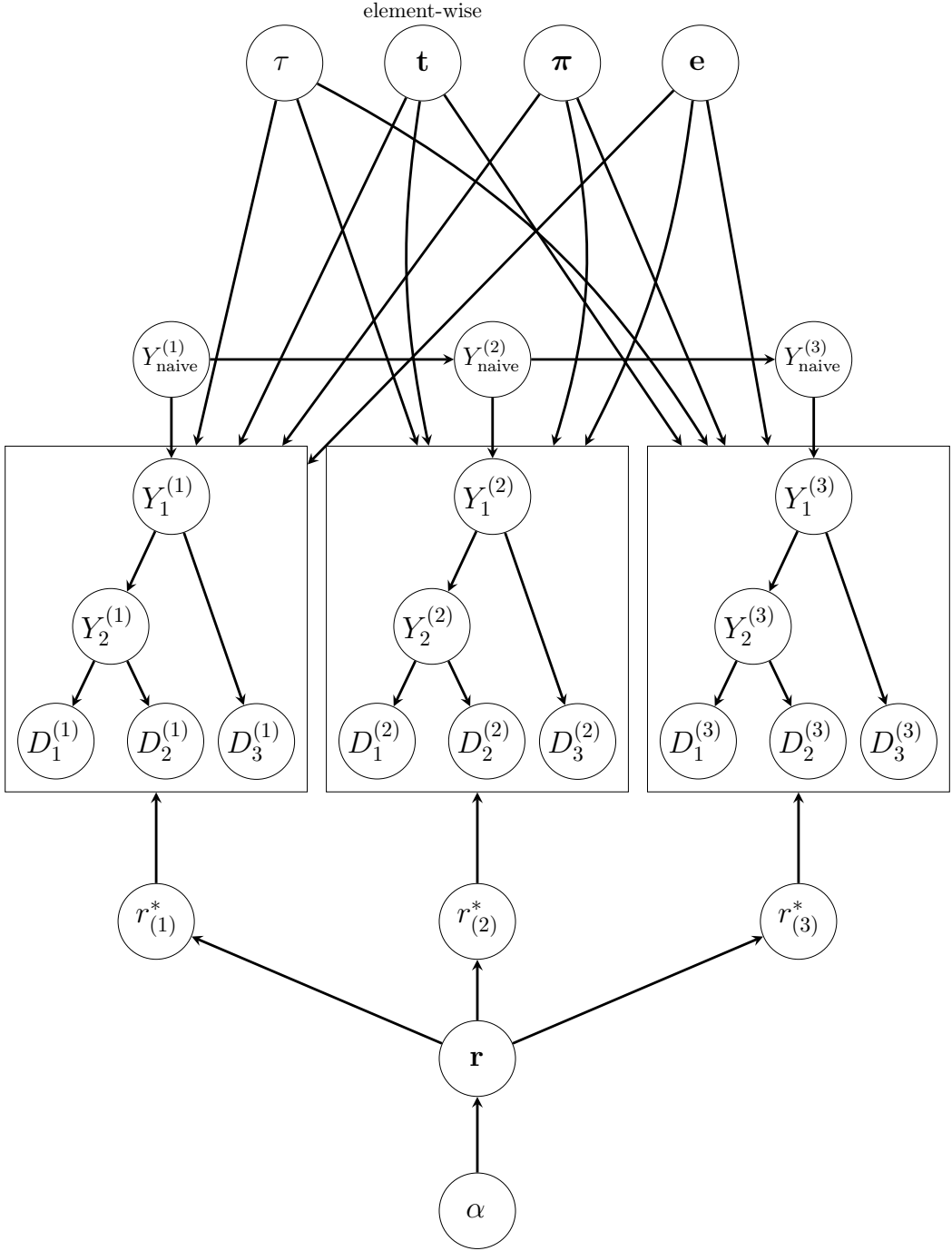

To help us describe the phylo-HMM generative process, we provide an illustration of the associated graphical model diagram for an example alignment with $m=3$ sequences and $n=3$ sites (\autoref*{phylo-hmm-dag}).
The naive sequence ``hidden state'' prior distribution $p(\mathbf{Y}_{\text{naive}})$ decomposes to $p\bigl(Y_{\text{naive}}^{(1)}\bigr) \prod_{j=2}^n p\bigl(Y_{\text{naive}}^{(j)} \, \big| \, Y_{\text{naive}}^{(j-1)}\bigr)$ and the bases are generated sequentially; these prior probabilities depend on hyperparameters that can be set using the \texttt{partis} software package \citep{ralph2016consistency, ralph2016likelihood}.
For the tree topology $\tau$, we assume that a tree is drawn from the Uniform distribution over $(m+1)$-tip unrooted trees; this seems like a strange choice given that we described $\tau$ as a rooted topology above, but this decision will be justified when we discuss how to perform Bayesian inference under the phylo-HMM.
The branch lengths $\mathbf{t}$ and the gamma shape parameter $\alpha$ are assumed to be \textit{a priori} independent and to follow $\Exponential(\lambda)$ distributions, where $\lambda$ is some prespecified rate.
The GTR exchangeability rates $\mathbf{e}$ and equilibrium base frequencies $\boldsymbol{\pi}$ are usually assumed to come from six-dimensional and four-dimensional Dirichlet distributions, respectively.

For each MSA site $j \in \{ 1, ..., n \}$, $r^*_{(j)}$ \textit{a priori} follows the $\Categorical\bigl(\mathbf{r}, \bigl(\frac{1}{K}, ..., \frac{1}{K}\bigr)\bigr)$ distribution.
Then, at each site $j \in \{ 1, ..., n \}$, we assume that $\mathbf{Y}_{\text{int}}^{(j)}$ and $\mathbf{D}^{(j)}$ are generated by drawing DNA states from CTMC transition probability matrices based on augmented branch lengths $\mathbf{t} \times r^*_{(j)}$.
For example, in \autoref*{phylo-hmm-dag}, we first sample $Y_1^{(j)}$ from $p\bigl(Y_1^{(j)} \, \big| \, Y_{\text{naive}}^{(j)}\bigr)$, which is a row vector distribution in the CTMC transition probability matrix for the ``branch length'' $t_1 \times r^*_{(j)}$, where $j \in \{ 1, ..., n \}$.
Once we have sampled $Y_1^{(j)}$ for $j = 1, ..., n$, we can draw $Y_2^{(j)}$ and $D_3^{(j)}$ using similar row vector distributions from CTMC transition probability matrices for the ``branch lengths'' $t_2 \times r^*_{(j)}$ and $t_3 \times r^*_{(j)}$, respectively.
We can recursively continue this process until we generate states at $D_1^{(j)}$ and $D_2^{(j)}$ for $j \in \{ 1, ..., n \}$.

\subsection*{Posterior Distribution Inference}

We are interested in sampling from the posterior distribution $p(\tau, \mathbf{t}, \boldsymbol{\pi}, \mathbf{e}, \alpha, \mathbf{r}, \mathbf{r}^*, \mathbf{Y}_{\text{naive}}, \mathbf{Y}_{\text{int}} \, | \, \mathbf{D})$:
\begin{align*}
&p(\tau, \mathbf{t}, \boldsymbol{\pi}, \mathbf{e}, \alpha, \mathbf{r}, \mathbf{r}^*, \mathbf{Y}_{\text{naive}}, \mathbf{Y}_{\text{int}} \, | \, \mathbf{D}) \\
&\propto p(\tau, \mathbf{t}, \boldsymbol{\pi}, \mathbf{e}, \alpha, \mathbf{r}, \mathbf{r}^*, \mathbf{Y}_{\text{naive}}, \mathbf{Y}_{\text{int}}, \mathbf{D}) \\
&=p(\mathbf{r}^*, \mathbf{Y}_{\text{int}}, \mathbf{D} \, | \, \tau, \mathbf{t}, \boldsymbol{\pi}, \mathbf{e}, \alpha, \mathbf{r}, \mathbf{Y}_{\text{naive}}) p(\tau) p(\mathbf{t}) p(\boldsymbol{\pi}) p(\mathbf{e}) p(\alpha) p(\mathbf{r} \, | \, \alpha) p\left(\mathbf{Y}_{\text{naive}}\right) \\
\begin{split}
&=\biggl\{ \prod_{j=1}^n p\Bigl(\mathbf{D}^{(j)} \, \Big| \, \tau, \mathbf{t}, \boldsymbol{\pi}, \mathbf{e}, r^*_{(j)}, \mathbf{Y}_{\text{int}}^{(j)}\Bigr) p\Bigl(\mathbf{Y}_{\text{int}}^{(j)} \, \Big| \, \tau, \mathbf{t}, \boldsymbol{\pi}, \mathbf{e}, r^*_{(j)}, Y_{\text{naive}}^{(j)}\Bigr) p\bigl(r^*_{(j)} \, \big| \, \mathbf{r}\bigr) \biggr\} \\
&\qquad \times p(\tau) p(\mathbf{t}) p(\boldsymbol{\pi}) p(\mathbf{e}) p(\alpha) p(\mathbf{r} \, | \, \alpha) p(\mathbf{Y}_{\text{naive}}),
\end{split}
\end{align*}
where this model decomposition results from the definition of a directed graphical model.
We also factorize the posterior distribution in the following way:
\begin{align*}
&p(\tau, \mathbf{t}, \boldsymbol{\pi}, \mathbf{e}, \alpha, \mathbf{r}, \mathbf{r}^*, \mathbf{Y}_{\text{naive}}, \mathbf{Y}_{\text{int}} \, | \, \mathbf{D}) \\
&\ =p(\tau, \mathbf{t}, \boldsymbol{\pi}, \mathbf{e}, \alpha, \mathbf{r} \, | \, \mathbf{D}) \\
&\qquad \times p(\mathbf{Y}_{\text{naive}} \, | \, \tau, \mathbf{t}, \boldsymbol{\pi}, \mathbf{e}, \alpha, \mathbf{r}, \mathbf{D}) \\
&\qquad \times p(\mathbf{r}^*, \mathbf{Y}_{\text{int}} \, | \, \tau, \mathbf{t}, \boldsymbol{\pi}, \mathbf{e}, \alpha, \mathbf{r}, \mathbf{Y}_{\text{naive}}, \mathbf{D}).
\end{align*}
This formulation is useful because it suggests that we can generate draws from the posterior distribution by sampling sequentially from three conditional probability distributions.
Conceptually, to sample from the posterior, we have to draw in-order: 1) the phylogeny-related parameters, 2) the naive sequence, and 3) the ancestral sequences.
We describe how to perform these three sampling steps in the following subsections and provide a complete summary of the sampling process in Algorithm \ref*{inf-summary}.

\begin{algorithm}[!ht]
\KwIn{CF multiple sequence alignment $\mathbf{D}$, number of discrete rates $K$\\ \ \ \quad \qquad $N_{\text{pool}}$, $N_{\text{final}}$ ($N_{\text{pool}} / N_{\text{final}} \approx 20$)}
\vspace{2mm}
\KwOut{$N_{\text{final}}$ samples of $(\tau, \mathbf{t}, \boldsymbol{\pi}, \mathbf{e}, \alpha, \mathbf{r}, \mathbf{r}^*, \mathbf{Y}_{\text{naive}}, \mathbf{Y}_{\text{int}}) \sim p(\tau, \mathbf{t}, \boldsymbol{\pi}, \mathbf{e}, \alpha, \mathbf{r}, \mathbf{r}^*, \mathbf{Y}_{\text{naive}}, \mathbf{Y}_{\text{int}} \, | \, \mathbf{D})$}\

\textit{Tree Sampling} --- $p(\tau, \mathbf{t}, \boldsymbol{\pi}, \mathbf{e}, \alpha, \mathbf{r} \, | \, \mathbf{D})$\
\vspace{3mm}
\begin{enumerate}[leftmargin=*]
\item Run \texttt{partis} on input data $\mathbf{D}$.
\end{enumerate}
\vspace{-2mm}
$\Rightarrow \mathbf{D}^* = \{ \mathbf{D}, \widehat{\mathbf{Y}}_{\text{naive}}^{\text{partis}} \}, \widehat{p}(\mathbf{Y}_{\text{naive}})$
\begin{enumerate}[leftmargin=*,start=2]
\item Run \texttt{RevBayes} MCMC on the augmented MSA $\mathbf{D}^*$.
\end{enumerate}
\vspace{-2mm}
$\Rightarrow N_{\text{pool}}\text{ samples of }(\tau, \mathbf{t}, \boldsymbol{\pi}, \mathbf{e}, \alpha, \mathbf{r}) \sim q(\tau, \mathbf{t}, \boldsymbol{\pi}, \mathbf{e}, \alpha, \mathbf{r} \, | \, \mathbf{D}^*)$
\begin{enumerate}[leftmargin=*,start=3]
\item Run the SIR algorithm without replacement on the $N_{\text{pool}}$ $(\tau, \mathbf{t}, \boldsymbol{\pi}, \mathbf{e}, \alpha, \mathbf{r})$ proposal samples with weights $w = \frac{p(\tau, \mathbf{t}, \boldsymbol{\pi}, \mathbf{e}, \alpha, \mathbf{r} \, | \, \mathbf{D})}{q(\tau, \mathbf{t}, \boldsymbol{\pi}, \mathbf{e}, \alpha, \mathbf{r} \, | \, \mathbf{D}^*)}$.
\end{enumerate}
\vspace{-2mm}
$\Rightarrow N_{\text{final}}\text{ samples of }(\tau, \mathbf{t}, \boldsymbol{\pi}, \mathbf{e}, \alpha, \mathbf{r}) \sim p(\tau, \mathbf{t}, \boldsymbol{\pi}, \mathbf{e}, \alpha, \mathbf{r} \, | \, \mathbf{D})$
\vspace{7.5mm}

\textit{Naive Sequence Sampling} --- $p(\mathbf{Y}_{\text{naive}} \, | \, \tau, \mathbf{t}, \boldsymbol{\pi}, \mathbf{e}, \alpha, \mathbf{r}, \mathbf{D})$\\[5mm]
For each sample $(\tau, \mathbf{t}, \boldsymbol{\pi}, \mathbf{e}, \alpha, \mathbf{r}) \sim p(\tau, \mathbf{t}, \boldsymbol{\pi}, \mathbf{e}, \alpha, \mathbf{r} \, | \, \mathbf{D})$:\\[2mm]
\hspace{25pt} For each site $j \in \{ n, ..., 1 \}$:
\begin{enumerate}[leftmargin=*]
\setlength\itemindent{50pt}
\item Draw $Y_{\text{naive}}^{(j)}$ using our phylo-HMM-based backward sampling procedure.
\end{enumerate}
\vspace{0mm}
$\Rightarrow N_{\text{final}}\text{ samples of }(\tau, \mathbf{t}, \boldsymbol{\pi}, \mathbf{e}, \alpha, \mathbf{r}, \mathbf{Y}_{\text{naive}}) \sim p(\tau, \mathbf{t}, \boldsymbol{\pi}, \mathbf{e}, \alpha, \mathbf{r}, \mathbf{Y}_{\text{naive}} \, | \, \mathbf{D})$
\vspace{7.5mm}

\textit{Intermediate Ancestral Sequence Sampling} --- $p(\mathbf{r}^*, \mathbf{Y}_{\text{int}} \, | \, \tau, \mathbf{t}, \boldsymbol{\pi}, \mathbf{e}, \alpha, \mathbf{r}, \mathbf{Y}_{\text{naive}}, \mathbf{D})$\\[5mm]
For each sample $(\tau, \mathbf{t}, \boldsymbol{\pi}, \mathbf{e}, \alpha, \mathbf{r}, \mathbf{Y}_{\text{naive}}) \sim p(\tau, \mathbf{t}, \boldsymbol{\pi}, \mathbf{e}, \alpha, \mathbf{r}, \mathbf{Y}_{\text{naive}} \, | \, \mathbf{D})$:\\[2mm]
\hspace{25pt} For each site $j \in \{ 1, ..., n \}$:
\begin{enumerate}[leftmargin=*]
\setlength\itemindent{50pt}
\item Sample $r^*_{(j)}$ according to probabilities proportional to $p\bigl(\mathbf{D}^{(j)} \, \big| \, \tau, \mathbf{t}, \boldsymbol{\pi}, \mathbf{e}, Y_{\text{naive}}^{(j)}, r^*_{(j)}\bigr)$.\\
\item Sample $\mathbf{Y}_{\text{int}}^{(j)}$ in a pre-order fashion using the standard ASR distribution at internal \phantom{\hspace{50pt}}nodes on a $r^*_{(j)}$-scaled tree.
\end{enumerate}
\vspace{0mm}
$\Rightarrow N_{\text{final}}\text{ samples of }(\tau, \mathbf{t}, \boldsymbol{\pi}, \mathbf{e}, \alpha, \mathbf{r}, \mathbf{r}^*, \mathbf{Y}_{\text{naive}}, \mathbf{Y}_{\text{int}}) \sim p(\tau, \mathbf{t}, \boldsymbol{\pi}, \mathbf{e}, \alpha, \mathbf{r}, \mathbf{r}^*, \mathbf{Y}_{\text{naive}}, \mathbf{Y}_{\text{int}} \, | \, \mathbf{D})$

\caption{Posterior Sampling of $p(\tau, \mathbf{t}, \boldsymbol{\pi}, \mathbf{e}, \alpha, \mathbf{r}, \mathbf{r}^*, \mathbf{Y}_{\text{naive}}, \mathbf{Y}_{\text{int}} \, | \, \mathbf{D})$}
\label{inf-summary}
\end{algorithm}

\subsubsection*{Tree Sampling}

Our strategy for sampling from $p(\tau, \mathbf{t}, \boldsymbol{\pi}, \mathbf{e}, \alpha, \mathbf{r} \, | \, \mathbf{D})$ is to first draw a large pool of observations from an easy-to-sample proposal distribution $q$ and then perform weighted bootstrap with weights $\frac{p}{q}$ on those samples to obtain approximate draws from the correct distribution.
This ``sampling-importance-resampling'' (SIR) algorithm is a sample filtering method that finds use in a wide variety of statistical applications \citep{smith1992bayesian,gordon1993novel,andrieu2010particle}.
The original SIR algorithm resampled observations with replacement, but there are theoretical and practical considerations that make resampling without replacement more attractive \citep{skare2003improved,gelman2013bayesian}.
A thorough review of SIR sampling can be found in \citep[Chapter 24]{gelman2004applied}.
We use the SIR algorithm to sample from $p(\tau, \mathbf{t}, \boldsymbol{\pi}, \mathbf{e}, \alpha, \mathbf{r} \, | \, \mathbf{D})$ because we want to take advantage of already-existing software programs for Bayesian phylogenetic inference while incorporating biologically-realistic details into our phylo-HMM.

Our phylogeny proposal distribution $q$ comes from the \texttt{RevBayes} software package \citep{hohna2016revbayes} because this is a proposal that is close to the target distribution $p$ while also being easy to sample from.
In short, this $q$ is traditional Bayesian phylogenetic analysis with a point estimate of the naive sequence.
In more detail, we first input the MSA $\mathbf{D}$ into the \texttt{partis} package, obtain a naive sequence point estimate $\widehat{\mathbf{Y}}_{\text{naive}}^{\text{partis}}$, and create an augmented MSA $\mathbf{D}^* = \{ \mathbf{D}, \widehat{\mathbf{Y}}_{\text{naive}}^{\text{partis}} \}$.
This allows us to generate Markov chain Monte Carlo (MCMC) samples from $q(\tau, \mathbf{t}, \boldsymbol{\pi}, \mathbf{e}, \alpha, \mathbf{r} \, | \, \mathbf{D}^*)$ and provides a convenient way to sample trees that have $m$ tips and an informative root branch length emanating from the naive sequence.
In our \texttt{RevBayes} analysis, we require that the prior components of $q(\tau, \mathbf{t}, \boldsymbol{\pi}, \mathbf{e}, \alpha, \mathbf{r} \, | \, \mathbf{D}^*)$ be defined as we discussed in the previous subsection for $p(\tau, \mathbf{t}, \boldsymbol{\pi}, \mathbf{e}, \alpha, \mathbf{r} \, | \, \mathbf{D})$ and as we will see, this assumption is critical to the validity of our technique.

For the purposes of this discussion, let us assume that we sample a large number (say $N_{\text{pool}}$) of parameter values from $q(\tau, \mathbf{t}, \boldsymbol{\pi}, \mathbf{e}, \alpha, \mathbf{r} \, | \, \mathbf{D}^*)$ in \texttt{RevBayes}.
As we briefly mentioned above, we resample $N_{\text{final}}$ times without replacement from the set of $N_{\text{pool}}$ $q(\tau, \mathbf{t}, \boldsymbol{\pi}, \mathbf{e}, \alpha, \mathbf{r} \, | \, \mathbf{D}^*)$ draws.
Each $q(\tau, \mathbf{t}, \boldsymbol{\pi}, \mathbf{e}, \alpha, \mathbf{r} \, | \, \mathbf{D}^*)$ sample is assigned a sampling weight $w = p(\tau, \mathbf{t}, \boldsymbol{\pi}, \mathbf{e}, \alpha, \mathbf{r} \, | \, \mathbf{D}) / q(\tau, \mathbf{t}, \boldsymbol{\pi}, \mathbf{e}, \alpha, \mathbf{r} \, | \, \mathbf{D}^*)$.
While it seems odd to use the ratio of posterior probabilities that are conditional on different datasets as bootstrap weights, the only technical requirement on $p$ and $q$ is that the parameters of interest (i.e.\ $\tau$, $\mathbf{t}$, $\boldsymbol{\pi}$, $\mathbf{e}$, $\alpha$, and $\mathbf{r}$) have the same support, which is indeed the case in our situation.
\citet{smith1992bayesian} suggest picking $N_{\text{pool}}$ and $N_{\text{final}}$ so $\frac{N_{\text{pool}}}{N_{\text{final}}} \geq 10$ while \citet{rubin1987calculation} proposed a safe rule-of-thumb to be $\frac{N_{\text{pool}}}{N_{\text{final}}} = 20$; we use $\frac{N_{\text{pool}}}{N_{\text{final}}} = 20$ in all the applied experiments conducted in this paper.
It may not be immediately clear how one would efficiently compute the numerator in $w$ so we express $w$ in the following form:
\begin{align*}
w &= \frac{p(\tau, \mathbf{t}, \boldsymbol{\pi}, \mathbf{e}, \alpha, \mathbf{r} \, | \, \mathbf{D})}{q(\tau, \mathbf{t}, \boldsymbol{\pi}, \mathbf{e}, \alpha, \mathbf{r} \, | \, \mathbf{D}^*)} \\
&= \frac{p(\tau, \mathbf{t}, \boldsymbol{\pi}, \mathbf{e}, \alpha, \mathbf{r}, \mathbf{D}) / p(\mathbf{D})}{q(\tau, \mathbf{t}, \boldsymbol{\pi}, \mathbf{e}, \alpha, \mathbf{r}, \mathbf{D}^*) / q(\mathbf{D}^*)} \\
&= \frac{p(\mathbf{D} \, | \, \tau, \mathbf{t}, \boldsymbol{\pi}, \mathbf{e}, \alpha, \mathbf{r})}{q(\mathbf{D}^* \, | \, \tau, \mathbf{t}, \boldsymbol{\pi}, \mathbf{e}, \alpha, \mathbf{r})} \frac{q(\mathbf{D}^*)}{p(\mathbf{D})} \\
&= \frac{\sum_{\mathbf{Y}_{\text{naive}}} p(\mathbf{D} \, | \, \tau, \mathbf{t}, \boldsymbol{\pi}, \mathbf{e}, \alpha, \mathbf{r}, \mathbf{Y}_{\text{naive}}) p(\mathbf{Y}_{\text{naive}} \, | \, \tau, \mathbf{t}, \boldsymbol{\pi}, \mathbf{e}, \alpha, \mathbf{r})}{q(\mathbf{D}^* \, | \, \tau, \mathbf{t}, \boldsymbol{\pi}, \mathbf{e}, \alpha, \mathbf{r})} \frac{q(\mathbf{D}^*)}{p(\mathbf{D})} \\
&= \frac{\sum_{\mathbf{Y}_{\text{naive}}} p(\mathbf{D} \, | \, \tau, \mathbf{t}, \boldsymbol{\pi}, \mathbf{e}, \mathbf{r}, \mathbf{Y}_{\text{naive}}) p(\mathbf{Y}_{\text{naive}})}{q(\mathbf{D}^* \, | \, \tau, \mathbf{t}, \boldsymbol{\pi}, \mathbf{e}, \alpha, \mathbf{r})} \frac{q(\mathbf{D}^*)}{p(\mathbf{D})} \\
&= \frac{\sum_{\mathbf{Y}_{\text{naive}}} \biggl\{ p\Bigl(Y_{\text{naive}}^{(1)}\Bigr) \prod_{j=2}^n p\Bigl(Y_{\text{naive}}^{(j)} \, \Big| \, Y_{\text{naive}}^{(j-1)}\Bigr) \prod_{k=1}^n p\Bigl(\mathbf{D}^{(k)} \, \Big| \, \tau, \mathbf{t}, \boldsymbol{\pi}, \mathbf{e}, \mathbf{r}, Y_{\text{naive}}^{(k)}\Bigr) \biggr\}}{q(\mathbf{D}^* \, | \, \tau, \mathbf{t}, \boldsymbol{\pi}, \mathbf{e}, \alpha, \mathbf{r})} \frac{q(\mathbf{D}^*)}{p(\mathbf{D})},
\end{align*}
where the transition from the second line to the third line is due to the priors on $\tau$, $\mathbf{t}$, $\boldsymbol{\pi}$, $\mathbf{e}$, $\alpha$, and $\mathbf{r}$ for both $p$ and $q$ being equal and the decomposition between the fourth and sixth lines is a result of $d$-separation conditional independencies \citep{lauritzen1996graphical} as follows.
Specifically, $\mathbf{Y}_{\text{naive}} \perp \{ \tau, \mathbf{t}, \boldsymbol{\pi}, \mathbf{e}, \alpha, \mathbf{r} \}$ because every undirected path in the graphical model between $\mathbf{Y}_{\text{naive}}$ and $\{ \tau, \mathbf{t}, \boldsymbol{\pi}, \mathbf{e}, \alpha, \mathbf{r} \}$ does not contain any ``conditioned'' child nodes and $\mathbf{D} \perp \alpha \, | \, \{ \tau, \mathbf{t}, \boldsymbol{\pi}, \mathbf{e}, \mathbf{r}, \mathbf{Y}_{\text{naive}} \}$ as all paths between $\mathbf{D}$ and $\alpha$ are ``blocked'' by the intermediate node $\mathbf{r}$.
The final denominator term above $q(\mathbf{D}^* \, | \, \tau, \mathbf{t}, \boldsymbol{\pi}, \mathbf{e}, \alpha, \mathbf{r})$ is the usual phylogenetic likelihood, here calculated by \texttt{RevBayes}.
Note that the marginal likelihood ratio $\frac{q(\mathbf{D}^*)}{p(\mathbf{D})}$ does not affect the bootstrap sampling because the sampling probabilities are proportional to $w$ and the likelihood ratio is a constant with respect to $\tau$, $\mathbf{t}$, $\boldsymbol{\pi}$, $\mathbf{e}$, $\alpha$, and $\mathbf{r}$ in the $N_{\text{pool}}$ sampling weights.
The numerator term in the final equation for $w$ looks complicated, but is actually the phylo-HMM likelihood with the ``hidden state'' probabilities $p\bigl(Y_{\text{naive}}^{(1)}\bigr)$ and $p\bigl(Y_{\text{naive}}^{(j)} \, \big| \, Y_{\text{naive}}^{(j-1)}\bigr)$ for $j \in \{ 2, ..., n \}$ and the ``emission probabilities'' $p\bigl(\mathbf{D}^{(k)} \, \big| \, \tau, \mathbf{t}, \boldsymbol{\pi}, \mathbf{e}, \mathbf{r}, Y_{\text{naive}}^{(k)}\bigr)$ that marginalize over the site-wise rates $r^*_{(k)}$ for $k \in \{ 1, ..., n \}$.
We can efficiently calculate the phylo-HMM likelihood using the forward algorithm \citep{rabiner1986introduction}, but this approach requires us to be able to compute the phylo-HMM emission probabilities in a straightforward manner.
The computation of the emission probabilities is of interest because the hidden state probabilities in $p(\mathbf{Y}_{\text{naive}})$ can be easily inferred using \texttt{partis}, which we now call $\widehat{p}(\mathbf{Y}_{\text{naive}})$.

It turns out that we can, again, leverage existing software tools to help us efficiently compute the emission probabilities $p\bigl(\mathbf{D}^{(k)} \, \big| \, \tau, \mathbf{t}, \boldsymbol{\pi}, \mathbf{e}, \mathbf{r}, Y_{\text{naive}}^{(k)}\bigr)$ for $k \in \{ 1, ..., n \}$.
The key point is to recognize that $p\bigl(\mathbf{D}^{(k)} \, \big| \, \tau, \mathbf{t}, \boldsymbol{\pi}, \mathbf{e}, \mathbf{r}, Y_{\text{naive}}^{(k)}\bigr)$ is an entry in the Felsenstein likelihood vector at the $Y_{\text{naive}}^{(k)}$ node, which denotes the probability of the observed data at only the tips that descend from node $Y_{\text{naive}}^{(k)}$, given the conditioned state of node $Y_{\text{naive}}^{(k)}$.
These vectors are commonly used within a post-order tree traversal algorithm to compute standard phylogenetic likelihoods \citep{felsenstein1981evolutionary}.
Let $\mathbf{F}_u = (F_{u1}, ..., F_{um})^T$ be the vector of partial likelihoods at node $u$, where $F_{ui}$ denotes the probability of the observed data at only the tips that descend from node $u$, given that the state of node $u$ is $i$.
Because we utilize an unrooted tree model in our phylo-HMM, we can use a Pulley Principle argument \citep{felsenstein1981evolutionary} to show that a standard phylogenetic likelihood on our tree can be represented as $p\bigl(\mathbf{D}^{(k)} \, \big| \, \tau, \mathbf{t}, \boldsymbol{\pi}, \mathbf{e}, \mathbf{r}, Y_{\text{naive}}^{(k)}\bigr) \pi_{Y_{\text{naive}}^{(k)}}$.
Thus, we can compute the phylo-HMM emission probabilities by first calculating the standard site-wise phylogenetic likelihoods on the same tree and then dividing out the $Y_{\text{naive}}^{(k)}$ stationary probabilities.
To compute these standard site-specific phylogenetic likelihoods, we make use of the \texttt{libpll} C library \citep{flouri2014phylogenetic}, which is a versatile high-performance software library for phylogenetic analysis.
Of course, we could have instead used a rooted tree model in our phylo-HMM, performed our own post-order tree traversal algorithm for likelihood computations, and extracted out the appropriate entries in the site-wise Felsenstein likelihood vectors, but we want our inference technique to scale to the large datasets that result from high-throughput BCR sequencing and no currently-existing software package meets this requirement.
In the next subsection, we describe how to generate naive sequence draws given the approximate phylogenetic tree samples from $p(\tau, \mathbf{t}, \boldsymbol{\pi}, \mathbf{e}, \alpha, \mathbf{r} \, | \, \mathbf{D})$.

\subsubsection*{Naive Sequence Sampling}
To sample naive sequences, we exploit the fact that our phylo-HMM is essentially a standard HMM with a naive-conditional phylogenetic likelihood as its emission distribution.
We draw the ``hidden state'' naive sequence $\mathbf{Y}_{\text{naive}}$ from $p(\mathbf{Y}_{\text{naive}} \, | \, \tau, \mathbf{t}, \boldsymbol{\pi}, \mathbf{e}, \alpha, \mathbf{r}, \mathbf{D})$ by adapting the hidden state posterior sampling technique for standard HMMs to be used for our specialized phylo-HMM \citep{scott2002bayesian}.
Just as we perform the forward algorithm by recursively computing and caching intermediate phylo-HMM likelihoods (i.e.\ forward probabilities) going left to right across the MSA, we can sample $\mathbf{Y}_{\text{naive}}$ by doing a backward pass through the phylo-HMM and drawing the $Y_{\text{naive}}^{(j)}$ states starting at site $n$ and ending at the first alignment site.
In fact, the maximum a posteriori probability (MAP) estimate of the hidden state sequence $\mathbf{Y}_{\text{naive}}$ is obtained using a similar procedure called the Viterbi algorithm \citep{rabiner1986introduction, scott2002bayesian}.
This backward sampling procedure can actually use the previously cached forward probabilities in the calculation of the sampling probabilities at each site, which is convenient because we already had to run the forward algorithm to sample the phylogeny-related parameters from $p(\tau, \mathbf{t}, \boldsymbol{\pi}, \mathbf{e}, \alpha, \mathbf{r} \, | \, \mathbf{D})$.
Once $\mathbf{Y}_{\text{naive}}$ has been sampled, we can proceed to draw the intermediate ancestral states $\mathbf{Y}_{\text{int}}$ from the conditional distribution $p(\mathbf{r}^*, \mathbf{Y}_{\text{int}} \, | \, \tau, \mathbf{t}, \boldsymbol{\pi}, \mathbf{e}, \alpha, \mathbf{r}, \mathbf{Y}_{\text{naive}}, \mathbf{D})$.

\subsubsection*{Intermediate Ancestral Sequence Sampling}

Just as we did for our naive sequence sampling, we sample the intermediate ancestral states $\mathbf{Y}_{\text{int}}$ by utilizing a modified version of the standard ancestral sequence reconstruction (ASR) technique used in phylogenetics \citep{nielsen2002mapping}.
It is important to note that sampling from $p(\mathbf{r}^*, \mathbf{Y}_{\text{int}} \, | \, \tau, \mathbf{t}, \boldsymbol{\pi}, \mathbf{e}, \alpha, \mathbf{r}, \mathbf{Y}_{\text{naive}}, \mathbf{D})$ can be reduced to drawing $\bigl(r^*_{(j)}, \mathbf{Y}_{\text{int}}^{(j)}\bigr)$ pairs from $p\bigl(r^*_{(j)}, \mathbf{Y}_{\text{int}}^{(j)} \, \big| \, \tau, \mathbf{t}, \boldsymbol{\pi}, \mathbf{e}, \alpha, \mathbf{r}, Y_{\text{naive}}^{(j)}, \mathbf{D}^{(j)}\bigr)$ for each MSA site $j \in \{ 1, ..., n \}$, which is justified by $d$-separation conditional independencies \citep{lauritzen1996graphical}.
At each site $j \in \{ 1, ..., n \}$, we first sample $r^*_{(j)}$ and then $\mathbf{Y}_{\text{int}}^{(j)}$ according to the previously-described distribution:
\begin{align*}
&p\Bigl(r^*_{(j)}, \mathbf{Y}_{\text{int}}^{(j)} \, \Big| \, \tau, \mathbf{t}, \boldsymbol{\pi}, \mathbf{e}, \alpha, \mathbf{r}, Y_{\text{naive}}^{(j)}, \mathbf{D}^{(j)}\Bigr) \\
&= p\Bigl(r^*_{(j)} \, \Big| \, \tau, \mathbf{t}, \boldsymbol{\pi}, \mathbf{e}, \alpha, \mathbf{r}, Y_{\text{naive}}^{(j)}, \mathbf{D}^{(j)}\Bigr) \, p\Bigl(\mathbf{Y}_{\text{int}}^{(j)} \, \Big| \, \tau, \mathbf{t}, \boldsymbol{\pi}, \mathbf{e}, \alpha, \mathbf{r}, Y_{\text{naive}}^{(j)}, r^*_{(j)}, \mathbf{D}^{(j)}\Bigr),
\end{align*}
where the above decomposition is based on the definition of conditional probability.
We draw the site-specific rates before the site-wise intermediate ancestral states because conditioning on the rates allows for simpler and more efficient ancestral state sampling using an already-existing software package.

It turns out that we can draw $r^*_{(j)}$ values from $\mathbf{r}$ with probabilities proportional to $p\bigl(\mathbf{D}^{(j)} \, | \, \tau, \mathbf{t}, \boldsymbol{\pi}, \mathbf{e}, Y_{\text{naive}}^{(j)}, r^*_{(j)}\bigr)$, which makes intuitive sense because it implies rates should be sampled according to the site-specific likelihoods with branch lengths scaled by the corresponding rates.
To understand why the above statement holds true, we express $p\bigl(r^*_{(j)} \, \big| \, \tau, \mathbf{t}, \boldsymbol{\pi}, \mathbf{e}, \alpha, \mathbf{r}, Y_{\text{naive}}^{(j)}, \mathbf{D}^{(j)}\bigr)$ as follows:
\begin{align*}
&p\Bigl(r^*_{(j)} \, \Big| \, \tau, \mathbf{t}, \boldsymbol{\pi}, \mathbf{e}, \alpha, \mathbf{r}, Y_{\text{naive}}^{(j)}, \mathbf{D}^{(j)}\Bigr) \\
&\propto p\Bigl(r^*_{(j)}, \mathbf{D}^{(j)} \, \Big| \, \tau, \mathbf{t}, \boldsymbol{\pi}, \mathbf{e}, \alpha, \mathbf{r}, Y_{\text{naive}}^{(j)}\Bigr) \\
&= p\Bigl(r^*_{(j)} \, \Big| \, \tau, \mathbf{t}, \boldsymbol{\pi}, \mathbf{e}, \alpha, \mathbf{r}, Y_{\text{naive}}^{(j)}\Bigr) \, p\Bigl(\mathbf{D}^{(j)} \, \Big| \, \tau, \mathbf{t}, \boldsymbol{\pi}, \mathbf{e}, \alpha, \mathbf{r}, Y_{\text{naive}}^{(j)}, r^*_{(j)}\Bigr) \\
&= p\Bigl(r^*_{(j)} \, \Big| \, \mathbf{r}\Bigr) \, p\Bigl(\mathbf{D}^{(j)} \, \Big| \, \tau, \mathbf{t}, \boldsymbol{\pi}, \mathbf{e}, Y_{\text{naive}}^{(j)}, r^*_{(j)}\Bigr) \\
&\propto p\Bigl(\mathbf{D}^{(j)} \, \Big| \, \tau, \mathbf{t}, \boldsymbol{\pi}, \mathbf{e}, Y_{\text{naive}}^{(j)}, r^*_{(j)}\Bigr),
\end{align*}
where the transition from the second line to the third line stems from the definition of conditional probability, the transition from the third to fourth line is a result of $d$-separation \citep{lauritzen1996graphical}, and the fourth-to-fifth line transition is due to the fact that $r^*_{(j)} \, | \, \mathbf{r} \sim \Categorical\bigl(\mathbf{r}, \bigl(\frac{1}{K}, ..., \frac{1}{K}\bigr)\bigr)$ by assumption.
These site-specific likelihoods are in fact almost identical to the naive-conditional phylogenetic likelihoods discussed in the previous two subsections with the only difference being that we now condition on the site-wise rates instead of marginalizing over them.

To illustrate why sampling $\mathbf{Y}_{\text{int}}^{(j)}$ from $p\bigl(\mathbf{Y}_{\text{int}}^{(j)} \, \big| \, \tau, \mathbf{t}, \boldsymbol{\pi}, \mathbf{e}, \alpha, \mathbf{r}, Y_{\text{naive}}^{(j)}, r^*_{(j)}, \mathbf{D}^{(j)}\bigr)$ is similar to drawing ASRs according to the procedure outlined by \citet{nielsen2002mapping}, we derive the sampling distribution of $Y_1^{(j)}$, the most recent common ancestor of $\mathbf{D}^{(j)}$.
The distribution $p\bigl(Y_1^{(j)} \, \big| \, \tau, \mathbf{t}, \boldsymbol{\pi}, \mathbf{e}, \alpha, \mathbf{r}, Y_{\text{naive}}^{(j)}, r^*_{(j)}, \mathbf{D}^{(j)}\bigr)$ can be decomposed in the following manner:
\begin{align*}
&p\Bigl(Y_1^{(j)} \, \Big| \, \tau, \mathbf{t}, \boldsymbol{\pi}, \mathbf{e}, \alpha, \mathbf{r}, Y_{\text{naive}}^{(j)}, r^*_{(j)}, \mathbf{D}^{(j)}\Bigr) \\
&\propto p\Bigl(Y_1^{(j)}, \mathbf{D}^{(j)} \, \Big| \, \tau, \mathbf{t}, \boldsymbol{\pi}, \mathbf{e}, \alpha, \mathbf{r}, Y_{\text{naive}}^{(j)}, r^*_{(j)}\Bigr) \\
&= p\Bigl(Y_1^{(j)} \, \Big| \, \tau, \mathbf{t}, \boldsymbol{\pi}, \mathbf{e}, \alpha, \mathbf{r}, Y_{\text{naive}}^{(j)}, r^*_{(j)}\Bigr) \, p\Bigl(\mathbf{D}^{(j)} \, \Big| \, \tau, \mathbf{t}, \boldsymbol{\pi}, \mathbf{e}, \alpha, \mathbf{r}, Y_{\text{naive}}^{(j)}, r^*_{(j)}, Y_1^{(j)}\Bigr) \\
&= p\Bigl(Y_1^{(j)} \, \Big| \, \tau, \mathbf{t}, \boldsymbol{\pi}, \mathbf{e}, Y_{\text{naive}}^{(j)}, r^*_{(j)}\Bigr) \, p\Bigl(\mathbf{D}^{(j)} \, \Big| \, \tau, \mathbf{t}, \boldsymbol{\pi}, \mathbf{e}, r^*_{(j)}, Y_1^{(j)}\Bigr),
\end{align*}
where, as before, the transition between the second and third lines are due to the definition of conditional probability and the transition from the third line to the fourth line is a result of $d$-separation \citep{lauritzen1996graphical}.
The first term in the resulting expression above, $p\bigl(Y_1^{(j)} \, \big| \, \tau, \mathbf{t}, \boldsymbol{\pi}, \mathbf{e}, Y_{\text{naive}}^{(j)}, r^*_{(j)}\bigr)$, is the CTMC transition probability between $Y_{\text{naive}}^{(j)}$ and $Y_1^{(j)}$ on a scaled branch length $t_1 \times r^*_{(j)}$ and the second term, $p\bigl(\mathbf{D}^{(j)} \, \big| \, \tau, \mathbf{t}, \boldsymbol{\pi}, \mathbf{e}, r^*_{(j)}, Y_1^{(j)}\bigr)$, is an entry in the Felsenstein likelihood vector at the $Y_1^{(j)}$ node on the given tree with $r^*_{(j)}$-scaled branch lengths.
This expression for the sampling distribution of $Y_1^{(j)}$ corresponds with the standard ASR distribution at internal nodes described in \citep{nielsen2002mapping}.
Thus, by conceptually treating $Y_{\text{naive}}^{(j)}$ as a sampled root node and scaling the branch lengths by $r^*_{(j)}$, we can draw $Y_1^{(j)} \sim p\bigl(Y_1^{(j)} \, \big| \, \tau, \mathbf{t}, \boldsymbol{\pi}, \mathbf{e}, \alpha, \mathbf{r}, Y_{\text{naive}}^{(j)}, r^*_{(j)}, \mathbf{D}^{(j)}\bigr)$ using equation (10) presented by \citet{nielsen2002mapping}.
By recursively using $d$-separation arguments \citep{lauritzen1996graphical}, we can use similar logic to that described above to sample states at the other internal nodes $\mathbf{Y}_{\text{int}}^{(j)} \setminus Y_1^{(j)}$ in a pre-order fashion as well.
Once we have drawn $\mathbf{Y}_{\text{int}}^{(j)}$ at each alignment site $j \in \{ 1, ..., n \}$, our three-stage sampling process of $p(\tau, \mathbf{t}, \boldsymbol{\pi}, \mathbf{e}, \alpha, \mathbf{r}, \mathbf{r}^*, \mathbf{Y}_{\text{naive}}, \mathbf{Y}_{\text{int}} \, | \, \mathbf{D})$ is completed.

\subsection*{Implementation}

The entire Bayesian phylo-HMM sampling process is implemented in a pipeline called \texttt{linearham}, which is available at \url{https://github.com/matsengrp/linearham}.
We developed our phylo-HMM data structure in C++ and, as mentioned before, used the tree structures from the \texttt{libpll} C library \citep{flouri2014phylogenetic} to ensure our phylogenetic likelihood computations were as fast as possible.
We have provided a Docker container (\url{https://hub.docker.com/r/matsengrp/linearham}) so users can try out the software without installation, as well as specifying the installation dependencies and provide an example command in a Dockerfile.

Our \texttt{linearham} program also provides an interface to \texttt{partis} to obtain $\widehat{p}(\mathbf{Y}_{\text{naive}})$.
It accepts a repertoire FASTA file as input and can infer the hidden state transition probabilities assuming either the repertoire needs to be partitioned into individual CFs before phylogenetic inference or the ``repertoire'' is a single CF already, which is useful for researchers that want to run \texttt{linearham} inference on a hand-curated CF.
It is also possible to specify an external set of $p(\mathbf{Y}_{\text{naive}})$ parameters.

In addition, \texttt{linearham} summarizes its phylogenetic inference output in a user-friendly format.
It provides an output FASTA file that contains each sampled amino acid naive sequence and its associated posterior probability, creates a FASTA-like file that maps each sampled amino acid naive sequence to its corresponding set of DNA naive sequences and posterior probabilities, and generates an amino acid naive sequence posterior probability logo using \texttt{WebLogo} \citep{crooks2004weblogo} to visualize the per-site uncertainties.
Furthermore, we provide similarly-styled output files for particular naive-to-tip tree lineages of interest by tabulating the posterior probabilities of sampled naive sequences and intermediate ancestral sequences on the lineage.
For naive-to-tip lineage analysis, we also create a posterior probability lineage graphic using \texttt{Graphviz} \citep{gansner2000open} that summarizes the different inferred naive-to-tip sequence trajectories and their relative confidences \citep{gong2013stability}.

\section*{Simulation Experiments}

In our simulation experiments, we focus on validating the accuracy of the naive sequence and ancestral sequence estimates produced by \texttt{linearham}.
For naive sequence validations, we compare and contrast the performance between \texttt{linearham}, \texttt{partis}, and the \texttt{ARPP} program \citep{kepler2013reconstructing}.
The \texttt{partis} package provides maximum likelihood naive sequence estimates \citep{ralph2016consistency}, but its model assumes a star-tree configuration, which is unrealistic for many CFs going through long periods of SHM and affinity maturation \citep{liao2013co,doria2014developmental,doria2016new}.
The \texttt{ARPP} program is also a likelihood-based framework that jointly infers a CF tree and the associated naive sequence using information about the V(D)J rearrangement process, but does not quantify phylogenetic or ancestral sequence uncertainty; we use this program in our validations because it is one of the only other programs that estimates CF phylogenies and naive sequences at the same time.
We did attempt to use the newer version of \texttt{ARPP} (called \texttt{Cloanalyst}), but ran into difficulties with the program crashing and were unable to successfully resolve this issue when we contacted the program author.

Similarly, for our ancestral sequence validations, we restrict our comparisons to the \texttt{linearham}, \texttt{RevBayes}, and \texttt{dnaml} \citep{phylip} packages.
As discussed before, the \texttt{RevBayes} program performs Bayesian phylogenetic inference on a given MSA, but in this case, we sample CF trees and ASRs from \texttt{RevBayes} using an augmented CF sequence alignment that contains the \texttt{partis}-inferred naive sequence (i.e.\ $\mathbf{D}^*$ from above).
This approach to ASR was used in \citep{simonich2019kappa} and is similar to that of \texttt{linearham} with the main difference being that \texttt{RevBayes} ASR sampling is conditional on the \texttt{partis}-inferred naive sequence whereas \texttt{linearham} ASR inference conditions on naive sequences drawn from a posterior distribution.
For all the experiments conducted in this section, we run \texttt{RevBayes} using 50,000 MCMC iterations, sampling every 10 iterations, discard the first 500 \texttt{RevBayes} samples as burn-in, and sample without replacement 225 times from the 4,500 effective \texttt{RevBayes} samples in the case of \texttt{linearham} inference.
The \texttt{dnaml} package performs maximum likelihood phylogenetic inference and generates an ASR conditional on this likelihood-based tree estimate.
While it is possible to sample ASRs on a maximum likelihood tree \citep{nielsen2002mapping}, \texttt{dnaml} only reports the most probable ancestral sequences.
We run \texttt{dnaml} on an augmented CF sequence alignment that contains the \texttt{ARPP} maximum likelihood estimate of the naive sequence.
In the following subsections, we describe our simulation experiments in more detail from the data-generating mechanism to the validation results.

\subsection*{Simulation Setup}

To simulate tree topologies with a fixed number of tips in our experiments, we used the single-parameter beta-splitting generative process \citep{aldous1996probability}.
The beta-splitting process is able to generate a wide variety of tree topologies ranging from balanced topologies (i.e.\ trees with approximately equal root-to-tip distances) to imbalanced topologies (i.e.\ trees with highly variable root-to-tip distances) by varying the associated ``balance'' parameter $\beta$.
Intuitively, this process can be seen as a recursive partitioning procedure that, at each tree split, partitions the $\Uniform(0, 1)$ random numbers corresponding to the ``tips'' on the current sub-interval according to a $\Beta(\beta + 1, \beta + 1)$ distribution.
As $\beta \rightarrow \infty$, the generated trees get closer and closer to balanced binary trees and, as $\beta \rightarrow -2$, the simulated topologies look more and more ``comb-like'' (i.e.\ imbalanced) \citep{aldous1996probability}.

We are motivated to use this topology-generating process because the level of balance of the tree determines the extent to which a phylogenetic approach to naive sequence estimation improves over a star-tree model.
Informally speaking, a phylogenetic approach weights the information coming from tips close to the root (in the imbalanced case) more strongly than tips more distant from the root.
Thus we expect a phylogenetic approach to be superior in the imbalanced case.
On the other hand, \texttt{partis} assumes evolution occurs according to a star tree, which implies the expected number of substitutions from the root to each of the tip sequences should be approximately equal.
Thus, for imbalanced trees, that have a large variance in the root-to-tip branch length distances, we would expect \texttt{partis} to provide poor naive sequence estimates for sequence datasets generated on those trees compared to a phylogenetic approach.
Throughout the rest of this section, we define ``tree imbalance'' to be the standard deviation of a tree's root-to-tip distances.

To generate branch lengths for our simulated trees that preserve the shapes induced by the beta-splitting topology prior, we independently draw values from a $\Uniform(0, 2M)$ distribution, where $M$ is a constant derived from HIV bnAb lineage trees.
Specifically, we ran the PC64 \citep{landais2017hiv} and VRC01 \citep{wu2015maturation} datasets through \texttt{partis} to obtain augmented CF sequence alignments with the \texttt{partis}-inferred naive sequence, inferred approximate maximum likelihood CF trees using \texttt{FastTree} \citep{price2009fasttree, price2010fasttree}, and set $M$ to be the average across all estimated branch lengths in the two trees ($\approx 0.0179$).
We describe these two datasets in more detail in the next section when we discuss our real-world dataset applications.
In the subsequent parts of this section, we refer to the inferred \texttt{FastTree} trees described above as the PC64 and VRC01 phylogenies, respectively.
We emphasize that these trees are only used as the basis for a simulation study and their level of accuracy is not especially important.

Each simulated phylogeny also receives a root branch length, which is not accounted for by the above generative processes.
We use the mean of the PC64 and VRC01 root branch lengths ($\approx 0.01759$) as the default root branch length for simulation and also assign simulated trees root branch lengths of $0.1$ to validate inference in settings with long periods of shared mutational history.
For each simulated tree, we draw a naive sequence at the root according to the \texttt{partis} prior distribution that, as mentioned previously, models the V(D)J rearrangement process.
We use the default settings in \texttt{partis} for human heavy chain data, since this represents the regime of most common interest.

To simulate DNA sequence data on the selected trees given the \texttt{partis}-generated naive sequences, we use the simulator in the \texttt{samm} package \citep{feng2019}.
A complex collection of enzymes introduces mutations to affinity-maturing sequences in a random pattern that is known to be highly sensitive to the sequence motif (i.e.\ the subsequence of DNA bases surrounding the mutating position) \citep{rogozin1992somatic, dunn1998base, chahwan2012aiding, methot2017molecular}.
The \texttt{samm} program estimates these motif mutabilities (i.e.\ the probability a position will mutate given the motif at that position) and substitution probabilities (i.e.\ the probability a position will mutate to a new base given the motif at that position) using a penalized Cox proportional hazards model and simulates sequence mutations given these inferred parameters.
In our work, we use the default \texttt{samm} parameters inferred for human heavy chain sequences \citep{feng2019}.

In our simulation experiments, we set the beta-splitting ``balance'' parameter $\beta$ to $-1.5$, $-1.25$, and $-1$; the CF sequence count $N_{\text{CF}}$ to $40$ and $80$; and the root branch length $t_0$ to $0.01759$ and $0.1$.
There are thus 12 different parameter combinations, and for each of these, we simulate 15 trees for a grand total of 180 simulated trees.
While 180 simulated trees does not seem like a large number of Monte Carlo replicates, we are forced to constrain the number of trees on which we perform inference because the \texttt{ARPP} program has to be run by hand via a GUI.
Using these parameter settings allows for the simulation of both balanced and imbalanced trees.

\subsection*{Naive Sequence Validations}

For each of the 180 simulated trees, we validate the accuracy of the naive sequence estimates generated from the \texttt{linearham}, \texttt{partis}, and \texttt{ARPP} programs by computing the hamming distances (i.e.\ the number of mismatched characters between two equal-length strings) between the estimates and the true naive sequence.
We did this for the inferred DNA sequences directly and also for those same sequences translated to amino acids after inference was complete.
The \texttt{partis} and \texttt{ARPP} packages provide naive sequence point estimates, but \texttt{linearham} samples naive sequences from a posterior distribution so we take the naive sequence with the highest posterior probability as the \texttt{linearham} ``point estimate''.

\begin{figure}[!ht]
\centering
\includegraphics[width=0.8\textwidth]{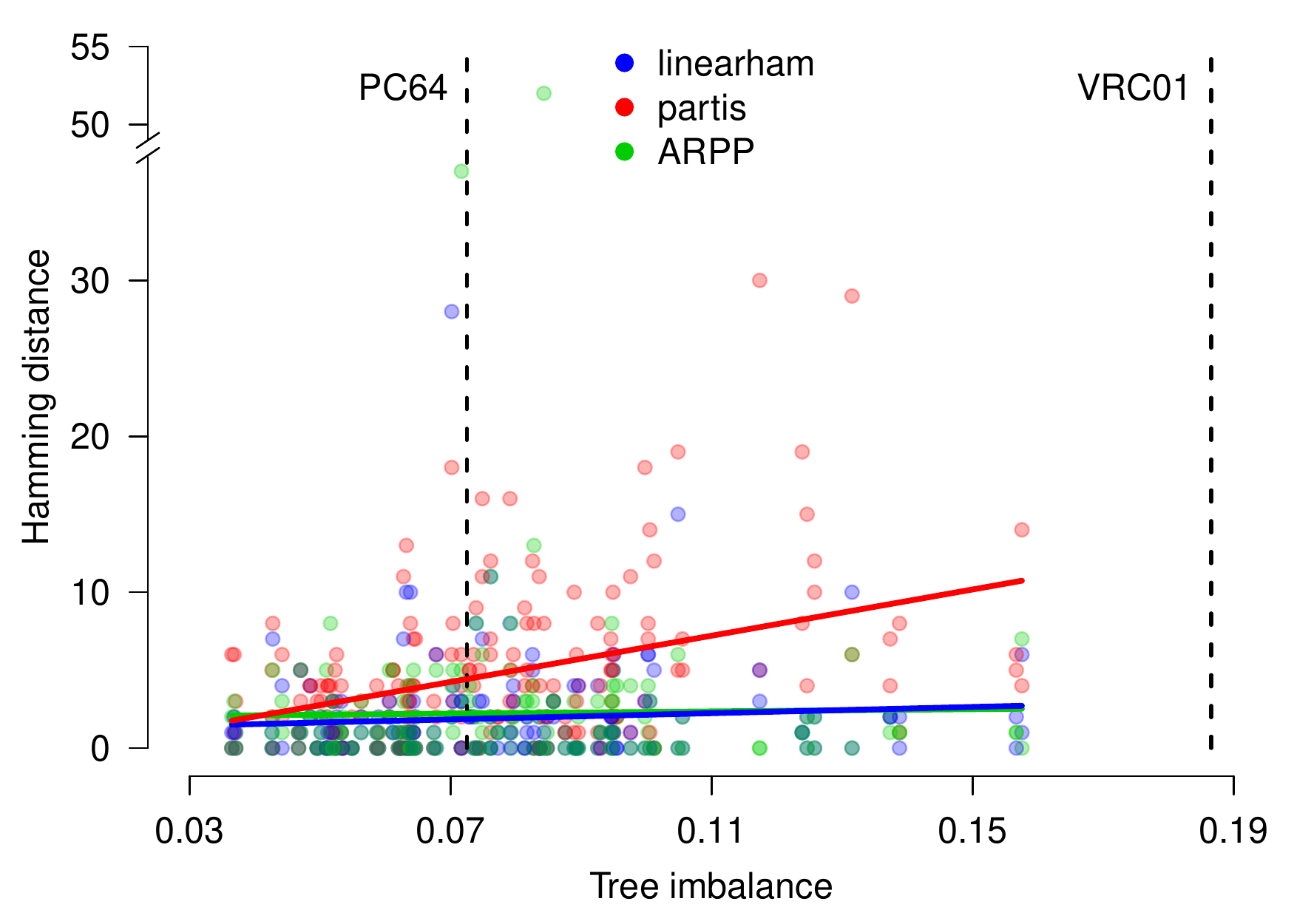}
\caption{The hamming distances between the simulated naive DNA sequences and their corresponding \texttt{linearham}, \texttt{partis}, and \texttt{ARPP} estimates versus the tree imbalance values of the simulated trees.
Linear regression lines are superimposed for each method to indicate how the results vary as trees get more imbalanced.  For reference, we plot the tree imbalance values for the PC64 and VRC01 trees.}
\label{naive_imbalance_plot}
\end{figure}

We summarize the hamming distance results for all 180 simulated trees described above and plot this performance metric against the corresponding values of tree imbalance (\autoref*{naive_imbalance_plot}); for reference, we plot the tree imbalance values for the PC64 and VRC01 trees as well.
The performance of \texttt{partis} clearly worsens as trees become more imbalanced, which makes sense given that \texttt{partis} assumes a star-tree configuration for clonal family evolution.
The \texttt{linearham} and \texttt{ARPP} programs provide accurate naive sequence estimates and perform similarly across the observed tree imbalance spectrum, which is not too surprising because they both incorporate phylogenetic information into their estimates.
We also show a summary of these results, split into values for the full-sequence and CDR3 regions, as well as the DNA/amino-acid sequence types (\autoref*{naive_imbalance_table}).
\autoref*{naive_imbalance_table} suggests that \texttt{linearham} and \texttt{ARPP} both perform better than \texttt{partis} does by approximately 2-3 nucleotides (1.6-1.8 amino acids) in the whole sequence and in the CDR3 region.
Furthermore, it seems \texttt{linearham} and \texttt{ARPP} perform similarly across all the different settings in \autoref*{naive_imbalance_table}.
\begin{table}[!ht]
\centering
\begin{tabular}{llcc}
Sequence Region & Program & \multicolumn{2}{c}{Sequence Type} \\
\hline
 & & DNA & Amino-Acid \\
\hline
\multirow{6}{*}{Full-sequence} & \multirow{2}{*}{\texttt{linearham}} & 1.92 & 1.17 \\
 & & (3.14) & (1.76) \\
 & \multirow{2}{*}{\texttt{partis}} & 4.81 & 3.02 \\
 & & (4.91) & (2.84) \\
 & \multirow{2}{*}{\texttt{ARPP}} & 2.24 & 1.39 \\
 & & (5.06) & (3.32) \\
\hline
\multirow{6}{*}{CDR3-only} & \multirow{2}{*}{\texttt{linearham}} & 1.66 & 1.08 \\
 & & (2.89) & (1.69) \\
 & \multirow{2}{*}{\texttt{partis}} & 4.17 & 2.69 \\
 & & (3.70) & (2.29) \\
 & \multirow{2}{*}{\texttt{ARPP}} & 1.27 & 0.856 \\
 & & (1.78) & (1.18) \\
\hline
\end{tabular}
\caption{Mean hamming distances between the simulated naive sequences and their corresponding estimates, where the hamming distances are averaged over all 180 simulated trees.  Results are provided for the \texttt{linearham}, \texttt{partis}, and \texttt{ARPP} programs; the full-sequence and CDR3 regions; and the DNA/amino-acid sequence types.  Standard errors are also presented in parentheses.}
\label{naive_imbalance_table}
\end{table}

We now present the mean hamming distance results, averaging over all trees generated under the different simulation parameter settings.
Specifically, we average over the trees that were simulated using beta-splitting ``balance'' parameter values $\beta = -1, -1.25, -1.5$, CF sequence counts $N_{\text{CF}} = 40, 80$, and root branch lengths $t_0 = 0.01759, 0.1$.
The mean hamming distance values seem to increase slightly for \texttt{linearham} and \texttt{ARPP} and considerably for \texttt{partis} as we go from $\beta = -1, -1.25$ to $\beta = -1.5$ (\autoref*{naive_imbalance_beta_table}), which is also suggested in \autoref*{naive_imbalance_plot}.
The performance of \texttt{linearham} and \texttt{partis} deteriorates as $N_{\text{CF}}$ goes from 40 to 80, while the opposite result is true for \texttt{ARPP} (\autoref*{naive_imbalance_nCF_table}).
Despite this, \texttt{linearham} is still better than \texttt{ARPP} at predicting the whole naive sequence.
As the root branch length $t_0$ increases from 0.01759 to 0.1, the mean hamming distances increase substantially for all three programs (\autoref*{naive_imbalance_t0_table}), which is intuitive because we are essentially introducing a higher number of mutations that are shared across all clonal sequences.
While the \texttt{linearham} naive sequence validations did not take advantage of the full naive sequence posterior distribution, we demonstrate the usefulness of accounting for phylogenetic uncertainty in our ancestral sequence validations described below.

\subsection*{Intermediate Ancestral Sequence Validations}

Our ancestral sequence validation experiments are centered around accurate inference of particular root-to-tip ancestral sequence lineages of interest because immunologists frequently use ASR to estimate the mutational pathways associated with antibody development \citep{doria2014developmental,simonich2019kappa}.
For each of our simulated trees, we determine the root-to-tip ancestral lineage of interest by identifying the tip sequence that is farthest from the naive sequence in terms of branch length distance.

We quantify the results of our ancestral sequence validation by treating it as a machine learning classification problem: do the posterior probabilities aid us in deciding if the ancestral lineage sequences are correct?
In all our experiments, we measure classification performance by recording the positive predictive value (i.e.\ the fraction of sequences in \textit{the ancestral lineage prediction set} that are \textit{on the true ancestral lineage}) and the true positive rate (i.e.\ the fraction of sequences \textit{on the true ancestral lineage} that are in \textit{the ancestral lineage prediction set}).
The ``predicted classification'' of these sequences is obtained by applying a decision boundary $\rho \in \{ 0.25, 0.5, 0.75 \}$ to the posterior probability.
Thus, for example, if $\rho = 0.75$ and a given ancestral sequence is on the true ancestral lineage and has posterior probability $0.8$, it is considered to be a ``true positive'' prediction.
This analysis is straightforward for the \texttt{linearham} and \texttt{RevBayes} programs, that do estimate posterior probabilities.
We define \texttt{dnaml} ``posterior probabilities'' to be either 0 or 1 depending on whether a lineage sequence is either outside or inside of the \texttt{dnaml}-inferred set of most probable reconstructed lineage sequences.

\begin{figure}[!ht]
\centering
\includegraphics[width=\textwidth]{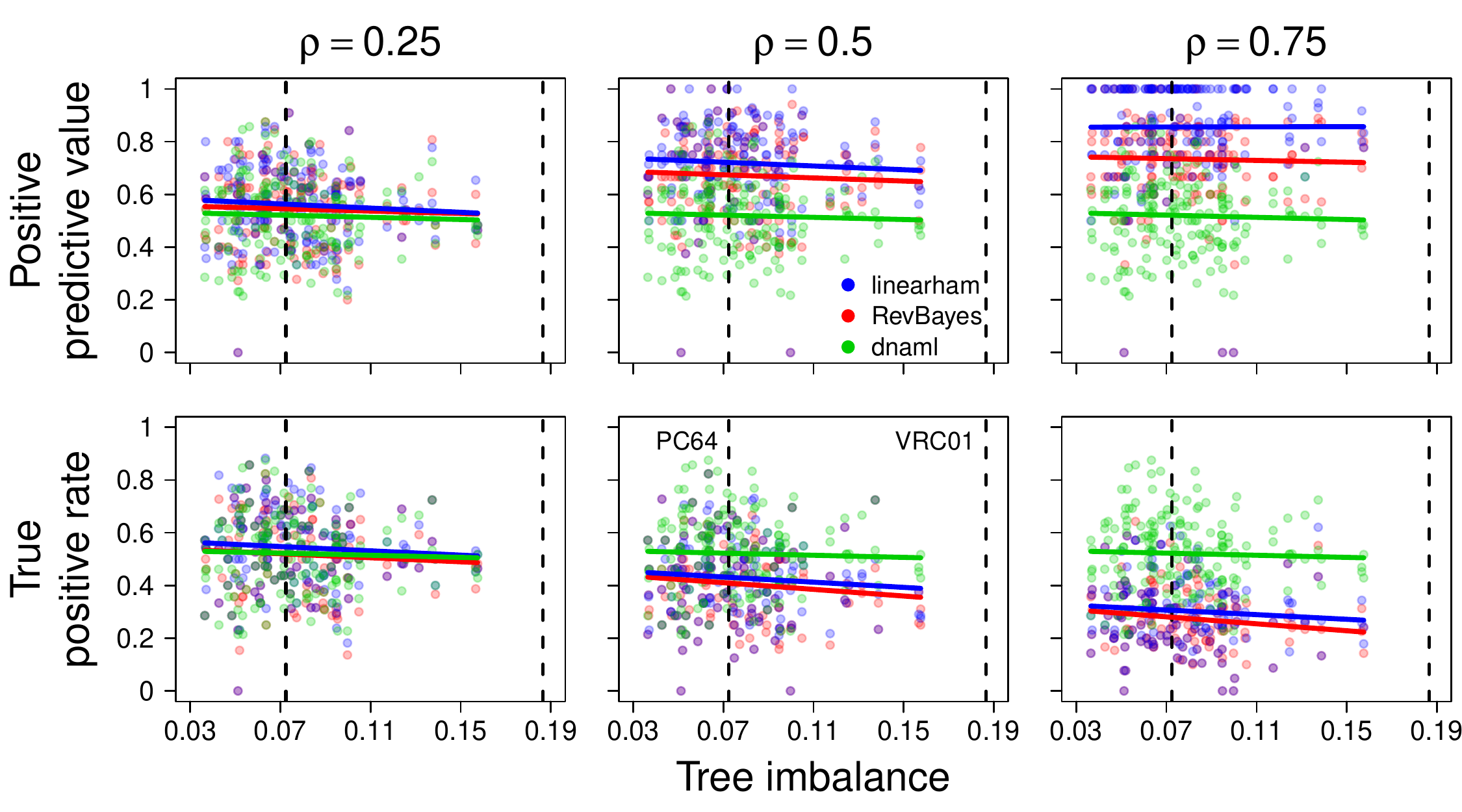}
\caption{The positive predictive values and the true positive rates versus the tree imbalance values of the simulated trees, stratified by decision boundary $\rho$.  Positive predictive values and true positive rates are computed on the DNA sequences and for the \texttt{linearham}, \texttt{RevBayes}, and \texttt{dnaml} programs.  Linear regression lines are superimposed for each package to indicate how the results vary as trees get more imbalanced.  For reference, we plot the tree imbalance values for the PC64 and VRC01 trees (vertical dashed lines).}
\label{asr_imbalance_plot}
\end{figure}

We report the positive predictive values and the true positive rates for all 180 simulated trees and plot these performance metrics against the corresponding values of tree imbalance (\autoref*{asr_imbalance_plot}); for reference, we plot the tree imbalance values for the PC64 and VRC01 trees as well.
Notice that the superimposed linear regression lines have negative slopes close to 0, which suggests that ancestral lineage inference is not clearly sensitive to the ``balance'' of the tree.
We also display the mean positive predictive values and mean true positive rates aggregated over all 180 trees for the same three programs, the different decision boundaries $\rho \in \{ 0.25, 0.5, 0.75 \}$, and the DNA/amino-acid sequence types (\autoref*{asr_imbalance_table}).
\autoref*{asr_imbalance_table} indicates that \texttt{linearham} performs better than \texttt{RevBayes} in every setting, indicating that accounting for naive rearrangement uncertainty in our posterior distribution rather than conditioning on the \texttt{partis}-inferred naive sequence leads to more accurate ancestral lineage sequence estimates.
At the lowest decision boundary $\rho = 0.25$, \texttt{linearham} obtains slightly better positive predictive values and true positive rates than either \texttt{RevBayes} or \texttt{dnaml} does (\autoref*{asr_imbalance_table}).
As the decision boundary $\rho$ increases, \texttt{linearham} and \texttt{RevBayes} achieve higher positive predictive values at the expense of lower true positive rates, which makes sense because high values of $\rho$ imply that only lineage sequences with high posterior probabilities are predicted to be on the true lineage.
In addition, note that the increases in positive predictive values are greater than the decreases in true positive rates for \texttt{linearham} as $\rho$ increases.
Of course, \texttt{dnaml} obtains the same positive predictive values and true positive rates regardless of $\rho$ because its ``posterior probabilities'' are either 0 or 1.

We also present the mean positive predictive values and mean true positive rates for decision boundary $\rho = 0.5$, averaging over all trees generated under the different simulation parameter settings.
The validation performance of all the programs seems to decline, albeit slightly, from $\beta = -1$ to $\beta = -1.5$ (\autoref*{asr_imbalance_beta_table}), which makes sense given the trends in \autoref*{asr_imbalance_plot}.
For the most part, mean positive predictive values and mean true positive rates increase for \texttt{linearham}, \texttt{RevBayes}, and \texttt{dnaml} as $N_{\text{CF}}$ goes from 40 to 80 (\autoref*{asr_imbalance_nCF_table}), which seems surprising but for larger CFs, the root-to-tip lineage becomes larger and may explain this observed pattern.
As the root branch length $t_0$ increases from 0.01759 to 0.1, the ancestral lineage validation results worsen considerably for all three programs (\autoref*{asr_imbalance_t0_table}), which is logical because, as we stated above, we are essentially introducing a higher number of mutations shared across all clonal sequences when we utilize longer root branch lengths in our simulations.

These results help demonstrate why Bayesian ancestral lineage inference should be favored over likelihood-based approaches to intermediate lineage inference as the Bayesian posterior probabilities quantify the uncertainty in our sequence estimates.
In a real-life experimental setting, the decision boundary $\rho$ should be chosen based on the desired level of positive predictive values or true positive rates and our analysis provides some insight into the mapping between $\rho$ and these classification metrics.
In practice, immunologists are probably more interested in controlling positive predictive values than true positive rates because synthesizing computationally-inferred lineage sequences and testing their binding and neutralization abilities is a laborious and expensive endeavor.
Thus, knowing the approximate fraction of inferred intermediate lineage sequences that are on the true lineage is of the utmost importance to an immunologist.
\begin{table}[!ht]
\makebox[\textwidth][c]{
\begin{tabular}{llccccccc}
Performance Metric & Program & \multicolumn{7}{c}{Sequence Type} \\
\hline
 & & \multicolumn{3}{c}{DNA} & & \multicolumn{3}{c}{Amino-Acid} \\
\cline{3-5} \cline{7-9}
 & & $\rho = 0.25$ & $\rho = 0.5$ & $\rho = 0.75$ & & $\rho = 0.25$ & $\rho = 0.5$ & $\rho = 0.75$ \\
\hline
\multirow{6}{*}{Positive predictive value} & \multirow{2}{*}{\texttt{linearham}} & 0.561 & 0.719 & 0.856 & & 0.613 & 0.758 & 0.858 \\
 & & (0.139) & (0.157) & (0.176) & & (0.131) & (0.145) & (0.146) \\
 & \multirow{2}{*}{\texttt{RevBayes}} & 0.544 & 0.672 & 0.734 & & 0.590 & 0.713 & 0.774 \\
 & & (0.141) & (0.160) & (0.166) & & (0.134) & (0.145) & (0.139) \\
 & \multirow{2}{*}{\texttt{dnaml}} & 0.520 & 0.520 & 0.520 & & 0.590 & 0.590 & 0.590 \\
 & & (0.139) & (0.139) & (0.139) & & (0.165) & (0.165) & (0.165) \\
 \hline
\multirow{6}{*}{True positive rate} & \multirow{2}{*}{\texttt{linearham}} & 0.545 & 0.428 & 0.304 & & 0.640 & 0.533 & 0.398 \\
 & & (0.146) & (0.147) & (0.125) & & (0.139) & (0.144) & (0.138) \\
 & \multirow{2}{*}{\texttt{RevBayes}} & 0.517 & 0.406 & 0.276 & & 0.606 & 0.505 & 0.370 \\
 & & (0.147) & (0.144) & (0.116) & & (0.140) & (0.144) & (0.134) \\
 & \multirow{2}{*}{\texttt{dnaml}} & 0.521 & 0.521 & 0.521 & & 0.584 & 0.584 & 0.584 \\
 & & (0.142) & (0.142) & (0.142) & & (0.166) & (0.166) & (0.166) \\
\hline
\end{tabular}
}
\caption{Mean positive predictive values and mean true positive rates, averaged over all 180 simulated trees.  Results are provided for the \texttt{linearham}, \texttt{RevBayes}, and \texttt{dnaml} programs; the different decision boundaries $\rho \in \{ 0.25, 0.5, 0.75 \}$; and the DNA/amino-acid sequence types.  Standard errors are also presented in parentheses.}
\label{asr_imbalance_table}
\end{table}

\section*{PC64/VRC01 Ancestral Lineage Analysis}

We illustrate the capabilities of \texttt{linearham} and \texttt{ARPP}/\texttt{dnaml} on real-world datasets by applying the three methods to subsets of the PC64 and VRC01 datasets.
The PC64 dataset contains a set of clonal sequences with multiple HIV-binding bNAbs that results from a longitudinal study over 46 months on an African donor (i.e.\ donor PC64) within the International AIDS Vaccine Initiative Protocol C cohort \citep{landais2016broadly}.
Our VRC01 clonal family dataset also originates from an HIV-infected donor and contains many bNAb sequences that are part of a well-known class of HIV-binding antibodies (i.e.\ the VRC01 class) \citep{wu2011focused, west2012structural, zhou2013multidonor, wu2015maturation}.
The tip sequences of interest for the PC64 and VRC01 datasets are chosen to be PCT64-35M and NIH45-46, respectively, which are both monoclonal antibody sequences that have accumulated a large amount of SHM.
We use 100 sequences from the PC64 CF dataset using a pruning strategy discussed in \citep{simonich2019kappa} and trim the VRC01 CF dataset to 268 sequences using the \texttt{cd-hit} sequence clustering program \citep{li2006cd} with a 95\% sequence identity cutoff.
We perform inference on these subsetted datasets using the same settings as described for our simulation experiments.

\begin{figure}[!ht]
\begin{subfigure}[!ht]{\textwidth}
\makebox[\textwidth][c]{
\includegraphics[width=1.2\textwidth]{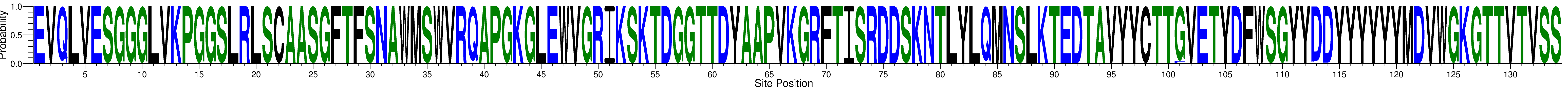}
}
\makebox[\textwidth][c]{
\includegraphics[width=1.2\textwidth]{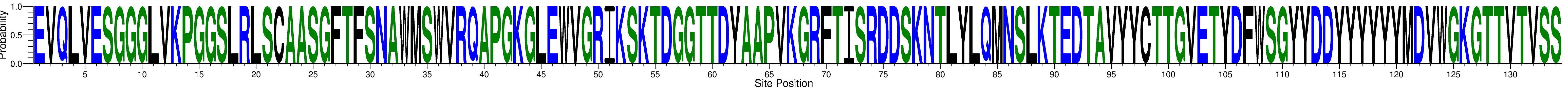}
}
\caption{PC64 naive sequence posterior probability logos}
\label{pc64_naive_logo}
\end{subfigure}
\newline
\vspace{5mm}
\newline
\begin{subfigure}[!ht]{\textwidth}
\makebox[\textwidth][c]{
\includegraphics[width=1.2\textwidth]{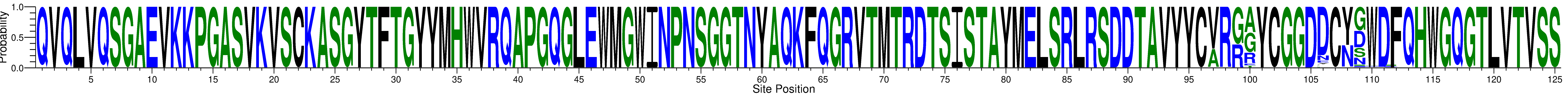}
}
\makebox[\textwidth][c]{
\includegraphics[width=1.2\textwidth]{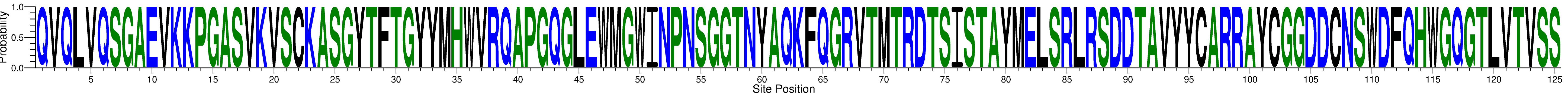}
}
\caption{VRC01 naive sequence posterior probability logos}
\label{vrc01_naive_logo}
\end{subfigure}
\caption{The \texttt{linearham}-inferred (top) and \texttt{ARPP}-inferred (bottom) amino acid naive sequence posterior probability logos for the pruned PC64 dataset of 100 sequences and the trimmed VRC01 alignment of 268 sequences.}
\label{naive_logos}
\end{figure}

The PC64 amino acid naive sequence posterior probability logos suggest that there is little uncertainty in the naive sequence reconstruction (\autoref*{pc64_naive_logo}) and, in fact, the most probable \texttt{linearham} amino acid naive sequence has a probability of approximately 0.92.
However, the VRC01 naive sequence seems to have considerable uncertainty in the CDR3 region (\autoref*{vrc01_naive_logo}), which shows that properly modeling ancestral sequence and phylogenetic uncertainty is important for real-world datasets with highly-mutated sequences.
The most probable \texttt{linearham} VRC01 naive sequence estimate has a posterior probability approximately equal to 0.036.
It is important to note that the VRC01 CF sequences were first collected 5 years after the diagnosis of the associated HIV-1 infection, whereas the PC64 CF sequences contain samples drawn from the corresponding donor as early as 4 months post infection.
This indicates that the VRC01 naive sequence reconstruction inherently has more uncertainty because there are not any early time-point samples in the corresponding dataset.
Furthermore, the most probable \texttt{linearham} VRC01 naive sequence amino acids do not perfectly match the corresponding \texttt{ARPP}-inferred residues (\autoref*{vrc01_naive_logo}).
In total, these results suggest that in the absence of uncertainty, \texttt{linearham} and \texttt{ARPP} produce similar naive sequence reconstructions.
However, when there is a significant amount of naive sequence uncertainty, \texttt{linearham}, unlike \texttt{ARPP}, provides alternate hypotheses that should be considered along with corresponding uncertainty estimates.

Our \texttt{linearham} analysis demonstrates that there are many probable naive-to-tip sequence paths (\autoref*{asr_lineages}), suggesting that intermediate ancestral sequences have high levels of uncertainty; we use 0.04 probability cutoffs in each \texttt{linearham}-inferred lineage graphic.
In particular, the \texttt{linearham} lineage diagram for the PC64 dataset (\autoref*{pc64_asr_lineage}) shows many possible routes of evolution from the different naive sequences to the PCT64-35M mature sequence and displays confidence values via posterior probabilities.
The VRC01 lineage graphic in \autoref*{vrc01_asr_lineage} does not display connections between any naive sequences and intermediate ancestral sequences, which reflects the fact that naive sequence inference is extremely challenging on this dataset.
In summary, there is considerable uncertainty in naive-to-tip mutational trajectory inference in real-world BCR datasets.
This finding contradicts the assertions of \citet{kepler2013reconstructing}, who states that ancestral sequence and phylogenetic uncertainty is unimportant, and proceeds with a single point estimate.
\begin{figure}[!ht]
\begin{subfigure}[!ht]{\textwidth}
\makebox[\textwidth][c]{
\minipage{0.4\textwidth}
\includegraphics[width=1.9\textwidth]{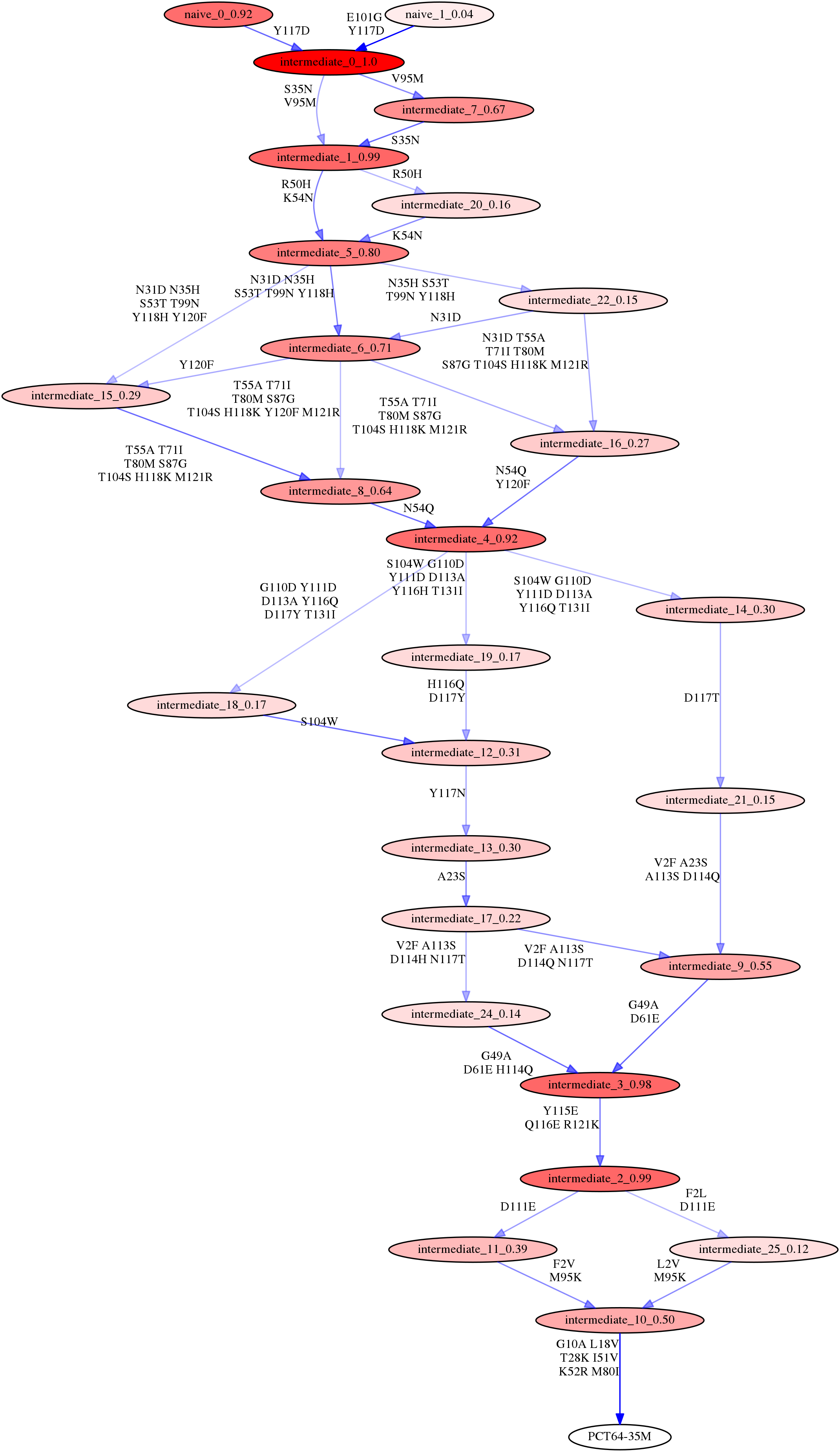}
\endminipage
\phantom{\hspace{2.5in}}
\minipage{0.4\textwidth}
\includegraphics[width=0.36\textwidth]{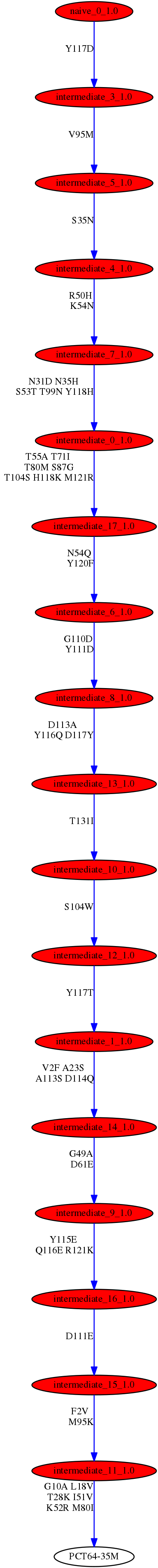}
\endminipage
}
\caption{PC64 posterior probability lineage inference}
\label{pc64_asr_lineage}
\end{subfigure}
\end{figure}
\begin{figure}[!ht]\ContinuedFloat
\begin{subfigure}[!ht]{\textwidth}
\makebox[\textwidth][c]{
\minipage{0.4\textwidth}
\includegraphics[width=1.75\textwidth]{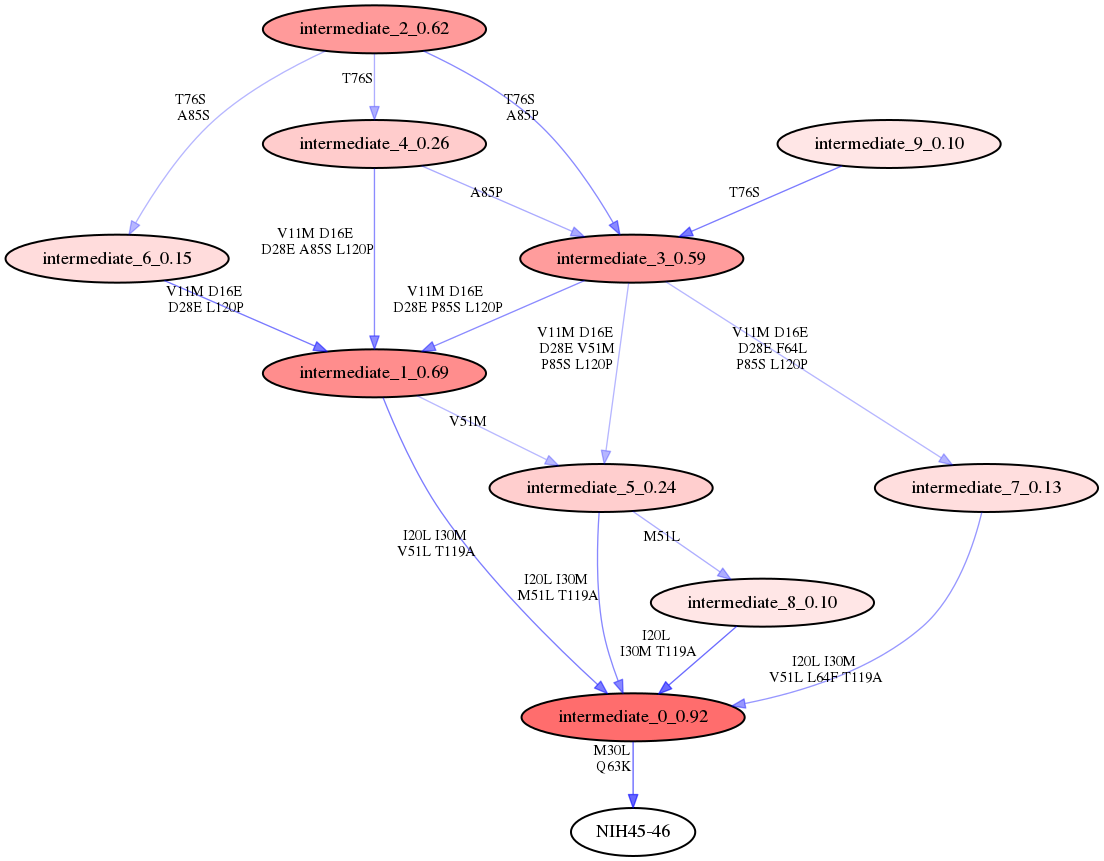}
\endminipage
\phantom{\hspace{2.25in}}
\minipage{0.4\textwidth}
\includegraphics[width=0.75\textwidth]{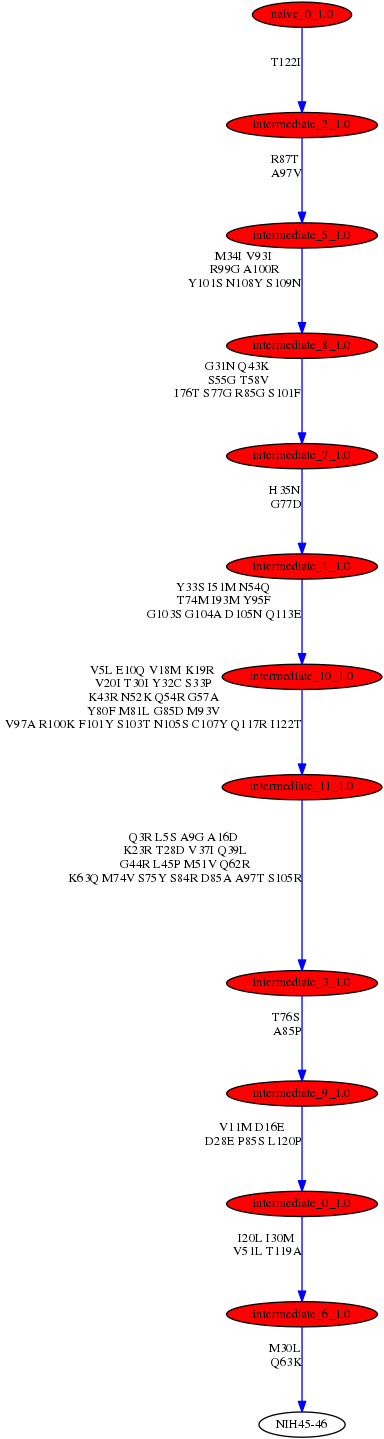}
\endminipage
}
\caption{VRC01 posterior probability lineage inference}
\label{vrc01_asr_lineage}
\end{subfigure}
\caption{The \texttt{linearham}-inferred (left) and \texttt{dnaml}-inferred (right) naive-to-tip amino acid sequence trajectories for the pruned PC64 dataset of 100 sequences and the trimmed VRC01 alignment of 268 sequences.  The tip sequences of interest for the PC64 and VRC01 datasets are chosen to be PCT64-35M and NIH45-46, respectively, and we use 0.04 probability cutoffs for these lineage graphics.  The nodes correspond to unique ancestral sequences filled with red color, where the opacity is proportional to the posterior probability of the associated sequence.  Each node has a label that denotes whether the associated sequence is a naive or intermediate ancestral sequence, the posterior probability rank of the sequence among all sampled naive or intermediate ancestral sequences, and the sequence-specific posterior probability itself.  The directed edges connecting nodes represent ancestral sequence transitions, are shaded blue with an opacity proportional to the posterior probability of the associated sequence transition, and are annotated with the site-specific mutations between the two sequences.}
\label{asr_lineages}
\end{figure}

\section*{Discussion}

In this paper, we introduce a novel Bayesian approach to CF phylogenetic inference that is based on a phylo-HMM.
Our phylo-HMM posterior sampling methodology not only allows for easy quantification of phylogenetic and ancestral sequence uncertainty but also models the V(D)J recombination process as an informative prior on the root sequence.
Specifically, our phylo-HMM models both the naive rearrangement and SHM processes by using a hidden state discrete-time Markov model on naive sequences that explicitly incorporates V(D)J rearrangement information and an emission distribution generating the clonal sequences conditional on the naive sequence that is based on phylogenetic likelihoods.
We show that our inference procedure, implemented in the software package \texttt{linearham}, provides higher-quality naive sequence and ancestral sequence estimates compared to those obtained under current state-of-the-art methods and augments these estimates with relative confidence values by reporting the associated posterior probabilities.

From our simulation experiments, we see that the \texttt{partis} naive sequence estimates get substantially worse as trees get more imbalanced.
This is in contrast to those of \texttt{linearham} and \texttt{ARPP}, which is intuitive because \texttt{partis} assumes a star-tree configuration whereas \texttt{linearham} and \texttt{ARPP} leverage phylogenetic models.
The \texttt{linearham} and \texttt{ARPP} programs perform similarly in our naive sequence validations regarding their most highly supported inferences, but only \texttt{linearham} is capable of characterizing naive sequence uncertainty.
This is important because, as we see in our VRC01 analysis (and in contrast to the findings of \citet{kepler2013reconstructing}), real datasets of significant practical importance can have very large uncertainties associated with their naive sequence estimates.
In addition, we demonstrate that \texttt{linearham} ancestral lineage inference performs better, via mean positive predictive values and mean true positive rates, than either \texttt{RevBayes} or \texttt{dnaml} does at the lowest decision boundary $\rho = 0.25$, which suggests that accounting for naive sequence, phylogenetic, and ancestral sequence uncertainty does lead to slightly improved ASR performance.
Furthermore, our Bayesian ASR results indicate that $\rho$ can be chosen according to a prespecified trade-off between positive predictive values and true positive rates, which is an important consideration for immunologists looking to synthesize computationally-inferred lineage sequences.

Based on all the evidence in this manuscript, we recommend using \texttt{linearham} to infer and visualize naive-to-mature mutational pathways in CFs harboring antibodies of interest.
Bayesian phylogenetic inference has already been shown to be useful for identifying different possible routes of evolution from a fixed naive sequence to a bNAb with relative confidence values \citep{simonich2019kappa} and we believe our Bayesian phylo-HMM analysis pipeline in \texttt{linearham} can be used to not only infer similarly-styled maturation pathways but also visualize the uncertainties inherent in the naive sequence and ancestral sequence estimates.
From a practical standpoint, these different possible ancestral lineages allow immunologists to generate many different intermediate antibody candidates that bind and/or neutralize HIV.

Our Bayesian phylo-HMM inference procedure admits a number of possible future extensions to enhance the effectiveness of our technique.
One drawback of our method is that it does not sample the parameters of $p(\mathbf{Y}_{\text{naive}})$ and uses \texttt{partis} (and its star-tree assumption) to estimate them.
Ideally, our Bayesian inference procedure would jointly sample all the model parameters, but currently this is not practically feasible.
In addition, our method implicitly assumes the CFs have been obtained from \texttt{partis}, which uses the star-tree assumption to cluster repertoire sequences.
It may be possible to, as \citet{ralph2016likelihood} state, incorporate the phylo-HMM in the CF clustering procedure within \texttt{partis} to obtain higher-quality CFs, but in our current Bayesian implementation that would be computationally costly.

\section*{Acknowledgments}

We thank Arman Bilge for many stimulating conversations about our phylo-HMM emission likelihood calculation, Andy Magee for answering all our questions about \texttt{RevBayes}, and Jean Feng for adapting the sequence simulator in \texttt{samm} to be used in our simulations.
We also want to thank Chaim Schramm for providing us with the VRC01 sequence dataset used in \citep{wu2015maturation}, Bryan Briney for sending us the PC64 multiple sequence alignment used in \citep{landais2017hiv}, Tyler Starr for creating the trimmed 268-sequence VRC01 dataset described above, and Laura Doepker for providing us with other experimental datasets to test \texttt{linearham} on.
This research was supported by NIH grants R01-GM113246, R01-AI120961, R01-AI138709, and U19-AI117891 as well as National Science Foundation grants CISE-1561334 and CISE-1564137.
The research of Frederick Matsen was supported in part by a Faculty Scholar grant from the Howard Hughes Medical Institute and the Simons Foundation.
Amrit Dhar was supported by an NSF IGERT DGE-1258485 fellowship.

\bibliographystyle{apa}
\bibliography{main.bib}

\beginsupplement
\clearpage

\pagenumbering{arabic}
\renewcommand*{\thepage}{S-\arabic{page}}

\renewcommand{\thesubsection}{S-\arabic{subsection}}

\section*{Supplementary Figures and Tables}

\begin{table}[!ht]
\makebox[\textwidth][c]{
\begin{tabular}{llccccccc}
Sequence Region & Program & \multicolumn{7}{c}{Sequence Type} \\
\hline
 & & \multicolumn{3}{c}{DNA} & & \multicolumn{3}{c}{Amino-Acid} \\
\cline{3-5} \cline{7-9}
 & & $\beta = -1$ & $\beta = -1.25$ & $\beta = -1.5$ & & $\beta = -1$ & $\beta = -1.25$ & $\beta = -1.5$ \\
\hline
\multirow{6}{*}{Full-sequence} & \multirow{2}{*}{\texttt{linearham}} & 1.73 & 1.92 & 2.10 & & 0.917 & 1.18 & 1.40 \\
 & & (3.94) & (2.66) & (2.71) & & (1.77) & (1.66) & (1.84) \\
 & \multirow{2}{*}{\texttt{partis}} & 3.85 & 3.82 & 6.77 & & 2.48 & 2.42 & 4.15 \\
 & & (3.26) & (3.85) & (6.48) & & (1.90) & (2.41) & (3.61) \\
 & \multirow{2}{*}{\texttt{ARPP}} & 2.42 & 2.52 & 1.78 & & 1.38 & 1.52 & 1.27 \\
 & & (6.89) & (5.11) & (1.98) & & (4.41) & (3.47) & (1.39) \\
\hline
\multirow{6}{*}{CDR3-only} & \multirow{2}{*}{\texttt{linearham}} & 1.45 & 1.67 & 1.87 & & 0.833 & 1.10 & 1.30 \\
 & & (3.42) & (2.55) & (2.66) & & (1.63) & (1.63) & (1.81) \\
 & \multirow{2}{*}{\texttt{partis}} & 3.53 & 3.47 & 5.50 & & 2.38 & 2.25 & 3.45 \\
 & & (2.84) & (3.45) & (4.34) & & (1.81) & (2.20) & (2.64) \\
 & \multirow{2}{*}{\texttt{ARPP}} & 1.12 & 1.32 & 1.38 & & 0.667 & 0.867 & 1.03 \\
 & & (1.76) & (1.95) & (1.63) & & (1.07) & (1.20) & (1.26) \\
\hline
\end{tabular}
}
\caption{Mean hamming distances between the simulated naive sequences and their corresponding estimates, where the hamming distances are averaged over all trees generated under the different beta-splitting ``balance'' parameter value settings.  Results are provided for the \texttt{linearham}, \texttt{partis}, and \texttt{ARPP} programs; the full-sequence and CDR3 regions; and the DNA/amino-acid sequence types.  Standard errors are also presented in parentheses.}
\label{naive_imbalance_beta_table}
\end{table}

\begin{table}[!ht]
\centering
\begin{tabular}{llccccccc}
Sequence Region & Program & \multicolumn{5}{c}{Sequence Type} \\
\hline
 & & \multicolumn{2}{c}{DNA} & & \multicolumn{2}{c}{Amino-Acid} \\
\cline{3-4} \cline{6-7}
 & & $N_{\text{CF}} = 40$ & $N_{\text{CF}} = 80$ & & $N_{\text{CF}} = 40$ & $N_{\text{CF}} = 80$ \\
\hline
\multirow{6}{*}{Full-sequence} & \multirow{2}{*}{\texttt{linearham}} & 1.73 & 2.10 & & 1.17 & 1.17 \\
 & & (2.36) & (3.77) & & (1.67) & (1.86) \\
 & \multirow{2}{*}{\texttt{partis}} & 4.12 & 5.50 & & 2.72 & 3.31 \\
 & & (3.58) & (5.90) & & (2.40) & (3.20) \\
 & \multirow{2}{*}{\texttt{ARPP}} & 2.39 & 2.09 & & 1.52 & 1.26 \\
 & & (4.36) & (5.70) & & (2.92) & (3.69) \\
\hline
\multirow{6}{*}{CDR3-only} & \multirow{2}{*}{\texttt{linearham}} & 1.51 & 1.81 & & 1.04 & 1.11 \\
 & & (2.26) & (3.41) & & (1.59) & (1.80) \\
 & \multirow{2}{*}{\texttt{partis}} & 3.84 & 4.49 & & 2.57 & 2.82 \\
 & & (3.38) & (3.98) & & (2.29) & (2.30) \\
 & \multirow{2}{*}{\texttt{ARPP}} & 1.33 & 1.21 & & 0.944 & 0.767 \\
 & & (1.63) & (1.92) & & (1.21) & (1.15) \\
\hline
\end{tabular}
\caption{Table analogous to \autoref*{naive_imbalance_beta_table}, but varying the CF sequence count $N_{\text{CF}}$.}
\label{naive_imbalance_nCF_table}
\end{table}

\begin{table}[!ht]
\centering
\begin{tabular}{llccccccc}
Sequence Region & Program & \multicolumn{5}{c}{Sequence Type} \\
\hline
 & & \multicolumn{2}{c}{DNA} & & \multicolumn{2}{c}{Amino-Acid} \\
\cline{3-4} \cline{6-7}
 & & $t_0 = 0.01759$ & $t_0 = 0.1$ & & $t_0 = 0.01759$ & $t_0 = 0.1$ \\
\hline
\multirow{6}{*}{Full-sequence} & \multirow{2}{*}{\texttt{linearham}} & 0.744 & 3.09 & & 0.456 & 1.88 \\
 & & (1.27) & (3.94) & & (0.889) & (2.10) \\
 & \multirow{2}{*}{\texttt{partis}} & 2.69 & 6.93 & & 1.77 & 4.27 \\
 & & (2.97) & (5.54) & & (1.91) & (3.06) \\
 & \multirow{2}{*}{\texttt{ARPP}} & 0.80 & 3.68 & & 0.467 & 2.31 \\
 & & (1.26) & (6.76) & & (0.737) & (4.46) \\
\hline
\multirow{6}{*}{CDR3-only} & \multirow{2}{*}{\texttt{linearham}} & 0.644 & 2.68 & & 0.433 & 1.72 \\
 & & (1.17) & (3.65) & & (0.875) & (2.04) \\
 & \multirow{2}{*}{\texttt{partis}} & 2.50 & 5.83 & & 1.68 & 3.71 \\
 & & (2.47) & (3.97) & & (1.62) & (2.42) \\
 & \multirow{2}{*}{\texttt{ARPP}} & 0.444 & 2.10 & & 0.311 & 1.40 \\
 & & (0.751) & (2.10) & & (0.574) & (1.37) \\
\hline
\end{tabular}
\caption{Table analogous to \autoref*{naive_imbalance_beta_table}, but varying the root branch length $t_0$.}
\label{naive_imbalance_t0_table}
\end{table}

\begin{table}[!ht]
\makebox[\textwidth][c]{
\begin{tabular}{llccccccc}
Performance Metric & Program & \multicolumn{7}{c}{Sequence Type} \\
\hline
 & & \multicolumn{3}{c}{DNA} & & \multicolumn{3}{c}{Amino-Acid} \\
\cline{3-5} \cline{7-9}
 & & $\beta = -1$ & $\beta = -1.25$ & $\beta = -1.5$ & & $\beta = -1$ & $\beta = -1.25$ & $\beta = -1.5$ \\
\hline
\multirow{6}{*}{Positive predictive value} & \multirow{2}{*}{\texttt{linearham}} & 0.742 & 0.715 & 0.700 & & 0.782 & 0.759 & 0.733 \\
 & & (0.157) & (0.157) & (0.158) & & (0.164) & (0.147) & (0.118) \\
 & \multirow{2}{*}{\texttt{RevBayes}} & 0.691 & 0.676 & 0.649 & & 0.725 & 0.726 & 0.689 \\
 & & (0.162) & (0.152) & (0.165) & & (0.166) & (0.140) & (0.124) \\
 & \multirow{2}{*}{\texttt{dnaml}} & 0.519 & 0.528 & 0.512 & & 0.575 & 0.605 & 0.590 \\
 & & (0.141) & (0.149) & (0.129) & & (0.179) & (0.173) & (0.141) \\
\hline
\multirow{6}{*}{True positive rate} & \multirow{2}{*}{\texttt{linearham}} & 0.435 & 0.443 & 0.407 & & 0.543 & 0.544 & 0.512 \\
 & & (0.149) & (0.139) & (0.152) & & (0.159) & (0.139) & (0.131) \\
 & \multirow{2}{*}{\texttt{RevBayes}} & 0.413 & 0.425 & 0.378 & & 0.510 & 0.522 & 0.482 \\
 & & (0.142) & (0.133) & (0.154) & & (0.160) & (0.136) & (0.133) \\
 & \multirow{2}{*}{\texttt{dnaml}} & 0.521 & 0.531 & 0.512 & & 0.566 & 0.600 & 0.586 \\
 & & (0.144) & (0.152) & (0.132) & & (0.182) & (0.166) & (0.148) \\
\hline
\end{tabular}
}
\caption{Mean positive predictive values and mean true positive rates for decision boundary $\rho = 0.5$, where we average over all trees generated under the different beta-splitting ``balance'' parameter value settings.  Results are provided for the \texttt{linearham}, \texttt{RevBayes}, and \texttt{dnaml} programs and the DNA/amino-acid sequence types.  Standard errors are also presented in parentheses.}
\label{asr_imbalance_beta_table}
\end{table}

\begin{table}[!ht]
\centering
\begin{tabular}{llccccccc}
Performance Metric & Program & \multicolumn{5}{c}{Sequence Type} \\
\hline
 & & \multicolumn{2}{c}{DNA} & & \multicolumn{2}{c}{Amino-Acid} \\
\cline{3-4} \cline{6-7}
 & & $N_{\text{CF}} = 40$ & $N_{\text{CF}} = 80$ & & $N_{\text{CF}} = 40$ & $N_{\text{CF}} = 80$ \\
\hline
\multirow{6}{*}{Positive predictive value} & \multirow{2}{*}{\texttt{linearham}} & 0.714 & 0.725 & & 0.759 & 0.756 \\
 & & (0.165) & (0.150) & & (0.166) & (0.121) \\
 & \multirow{2}{*}{\texttt{RevBayes}} & 0.654 & 0.690 & & 0.709 & 0.717 \\
 & & (0.171) & (0.147) & & (0.168) & (0.117) \\
 & \multirow{2}{*}{\texttt{dnaml}} & 0.513 & 0.526 & & 0.586 & 0.594 \\
 & & (0.146) & (0.132) & & (0.177) & (0.153) \\
\hline
\multirow{6}{*}{True positive rate} & \multirow{2}{*}{\texttt{linearham}} & 0.430 & 0.427 & & 0.527 & 0.539 \\
 & & (0.153) & (0.141) & & (0.159) & (0.127) \\
 & \multirow{2}{*}{\texttt{RevBayes}} & 0.403 & 0.408 & & 0.501 & 0.509 \\
 & & (0.151) & (0.137) & & (0.158) & (0.129) \\
 & \multirow{2}{*}{\texttt{dnaml}} & 0.514 & 0.528 & & 0.573 & 0.595 \\
 & & (0.152) & (0.132) & & (0.174) & (0.156) \\
\hline
\end{tabular}
\caption{Table analogous to \autoref*{asr_imbalance_beta_table}, but varying the CF sequence count $N_{\text{CF}}$.}
\label{asr_imbalance_nCF_table}
\end{table}

\begin{table}[!ht]
\centering
\begin{tabular}{llccccccc}
Performance Metric & Program & \multicolumn{5}{c}{Sequence Type} \\
\hline
 & & \multicolumn{2}{c}{DNA} & & \multicolumn{2}{c}{Amino-Acid} \\
\cline{3-4} \cline{6-7}
 & & $t_0 = 0.01759$ & $t_0 = 0.1$ & & $t_0 = 0.01759$ & $t_0 = 0.1$ \\
\hline
\multirow{6}{*}{Positive predictive value} & \multirow{2}{*}{\texttt{linearham}} & 0.728 & 0.711 & & 0.768 & 0.748 \\
 & & (0.141) & (0.173) & & (0.140) & (0.150) \\
 & \multirow{2}{*}{\texttt{RevBayes}} & 0.682 & 0.662 & & 0.727 & 0.700 \\
 & & (0.146) & (0.173) & & (0.135) & (0.153) \\
 & \multirow{2}{*}{\texttt{dnaml}} & 0.537 & 0.503 & & 0.616 & 0.564 \\
 & & (0.133) & (0.144) & & (0.154) & (0.172) \\
\hline
\multirow{6}{*}{True positive rate} & \multirow{2}{*}{\texttt{linearham}} & 0.450 & 0.407 & & 0.560 & 0.506 \\
 & & (0.136) & (0.154) & & (0.140) & (0.143) \\
 & \multirow{2}{*}{\texttt{RevBayes}} & 0.415 & 0.396 & & 0.523 & 0.487 \\
 & & (0.131) & (0.155) & & (0.135) & (0.151) \\
 & \multirow{2}{*}{\texttt{dnaml}} & 0.536 & 0.506 & & 0.609 & 0.559 \\
 & & (0.136) & (0.147) & & (0.157) & (0.171) \\
\hline
\end{tabular}
\caption{Table analogous to \autoref*{asr_imbalance_beta_table}, but varying the root branch length $t_0$.}
\label{asr_imbalance_t0_table}
\end{table}

\end{document}